\documentclass[usenatbib]{mn2e}

\usepackage{graphicx}
\usepackage{mathptmx}
\usepackage{subfigure}
\usepackage{amssymb}
\graphicspath{{./Images/}}
\begin{document}

\title[Synchrotron luminosity at $0<z<2$]{The relationship between
  star formation rate and radio synchrotron luminosity at $0<z<2$}

\author[T.\ Garn et al.]{Timothy Garn$^{1}$\thanks{E-mail: tsg@roe.ac.uk},
David A.\ Green$^{2}$, Julia M.\ Riley$^{2}$, Paul Alexander$^{2,3}$\\ 
    $^{1}$SUPA, Institute for Astronomy, Royal Observatory Edinburgh,
    Blackford Hill, Edinburgh EH9~3HJ\\
    $^{2}$Astrophysics Group, Cavendish Laboratory, 19 J.~J.~Thomson
    Ave., Cambridge CB3~0HE\\
    $^{3}$Kavli Institute of Cosmology Cambridge, Madingley Road,
    Cambridge, CB3 0HA} 

\date{\today}
\pagerange{\pageref{firstpage}--\pageref{lastpage}; } \pubyear{2009}
\label{firstpage}
\maketitle

%%%%%%%%%%%%%%%%%%%%%%%%%%%%%%%%%%%%%%%%%%%%%%%%%%
\begin{abstract}
We probe the relationship between star formation rate (SFR) and radio
synchrotron luminosity in galaxies at $0<z<2$ within the northern {\it
Spitzer} Wide-area Infrared Extragalactic survey (SWIRE) fields, in
order to investigate some of the assumptions that go into calculating
the star formation history of the Universe from deep radio
observations.  We present new 610-MHz Giant Metrewave Radio Telescope
(GMRT) observations of the European Large-Area {\it ISO} Survey
(ELAIS)-N2 field, and using this data, along with previous GMRT
surveys carried out in the ELAIS-N1 and Lockman Hole regions, we
construct a sample of galaxies which have redshift and SFR information
available from the SWIRE survey.  We test whether the local
relationship between SFR and radio luminosity is applicable to $z=2$
galaxies, and look for evolution in this relationship with both
redshift and SFR in order to examine whether the physical processes
which lead to synchrotron radiation have remained the same since the
peak of star formation in the Universe.  We find that the local
calibration between radio luminosity and star formation can be
successfully applied to radio-selected high-redshift, high-SFR
galaxies, although we identify a small number of sources where this
may not be the case; these sources show evidence for inaccurate
estimations of their SFR, but there may also be some contribution from
physical effects such as the recent onset of starburst activity, or
suppression of the radio luminosity within these galaxies.
\end{abstract}


%%%%%%%%%%%%%%%%%%%%%%%%%%%%%%%%%%%%%%%%%%%%%%%%%%
\section{Introduction}
\label{sec:introduction}
In recent years there has been much interest in constraining the star
formation history of the Universe
\citep[e.g.][]{Lilly96,Madau96,Mobasher99,Haarsma00,Hopkins03SDSS,Hopkins04,Hopkins06,Tresse07,Seymour08,Smolcic09SFR,Magnelli09}.
Estimates of the variation in the co-moving star formation rate
density (SFRD) of the Universe with redshift can be made from a number
of multi-wavelength tracers, with the most common ones being in the
optical and infrared regimes \citep[see][for a review of star
formation indicators]{Kennicutt98}.  These estimates can differ by up
to an order of magnitude \citep[e.g.][]{Hopkins03SDSS}, which is in
part due to the significant uncertainty in the correction factor that
needs to be applied for the effects of interstellar dust grains.
These grains absorb short-wavelength ultra-violet (UV) and optical
radiation, heat up to temperatures of a few tens to hundreds of
kelvin, and re-emit thermally with a modified black-body spectrum that
peaks in the infrared, thus having a significant effect on the optical
and infrared spectral energy distribution (SED) of a galaxy.

There have been a few attempts to estimate the SFRD of the Universe
from radio observations
\citep[e.g.][]{Mobasher99,Haarsma00,Seymour08,Smolcic09SFR}.  These
studies have a significant advantage over optical and infrared
estimates, in that radio observations are unaffected by the presence
of dust grains, and do not require the corrections that are needed at
other wavelengths.  However, there are a number of disadvantages to
radio studies, the principal one being that very deep observations are
required in order to reach the regime where star-forming galaxies
begin to dominate \citep[around 100~$\mu$Jy at 1.4~GHz,
e.g.][]{Seymour08}.  This disadvantage will be overcome with the next
generation of radio telescopes such as the Low Frequency Array (LOFAR)
and the Square Kilometre Array (SKA), which are expected to routinely
reach observation depths where the radio sky is dominated by
star-forming galaxies.

The non-thermal radio emission from star-forming galaxies is much
greater than the thermal emission at frequencies below a few GHz, and
is predominantly made up of synchrotron radiation, with supernova
remnants putting energy into a population of electrons which are
accelerated through the galactic magnetic field \citep[see
e.g.][]{Condon92}.  The details of this process are still a subject
for debate, with two principal alternatives; that galaxies are
optically thick to cosmic rays and act as `calorimeters'
\citep{Volk89}, or that galaxies are optically thin
\citep{Chi90,Helou93}, and the cosmic rays can diffuse out of the host
galaxy.  In the former case, the radio flux from a galaxy will be
essentially independent of the magnetic field $B$
\citep[e.g.][]{Thompson06}, since all of the energy from the cosmic
rays will be radiated within the galaxy.  In the latter case only some
fraction of the total energy will be radiated within the galaxy, and
variations in the $B$-field strength of galaxies over time may show up
as a variation in the confinement of cosmic rays, and an evolution in
the relationship between star formation rate (SFR) and synchrotron
luminosity with redshift.  Star-forming galaxies with magnetic field
strengths of a few $\mu$G have been indirectly detected through
Faraday rotation measurements out to $z\sim2$
\citep{Kronberg08,Bernet08}, but the rate and processes which lead to
evolution in $B$-field strength with time are currently unknown.

Attempts to estimate the SFRD of the Universe from radio observations
currently rely upon the untested assumption that local relationships
between radio luminosity and SFR \citep[e.g.][]{Condon90,Bell03} can
be applied successfully at higher redshift, and to sources undergoing
much more extreme starbursts than are seen in the local Universe.  In
this work we test these assumptions, using new radio surveys of the
three northern {\it Spitzer} Wide-area Infrared Extragalactic survey
\citep[SWIRE;][]{Lonsdale03} fields.  SWIRE is a 49~deg$^{2}$ study of
six regions of sky that have particularly low values of background
cirrus emission, using the {\it Spitzer Space Telescope}
\citep{Werner04}.  Observations have been made in seven {\it Spitzer}
wavebands with the Multiband Imaging Array for {\it Spitzer}
\citep[MIPS;][]{Rieke04} and the Infrared Array Camera
\citep[IRAC;][]{Fazio04} between 3.6~$\mu$m and 160~$\mu$m.  The three
SWIRE fields studied in this work are in the northern hemisphere --
the European Large Area {\it ISO} Survey-North~1 and -North~2 regions
(ELAIS-N1 and ELAIS-N2), and the Lockman Hole.  A great deal of
complementary data has been taken on the SWIRE fields; a band-merged
catalogue containing over one~million sources has been created
\citep[Data Release 2;][]{Surace05}, and a photometric redshift
catalogue has been generated \citep{RowanRobinson08} from the
multi-wavelength data.  All three northern regions are covered by the
Faint Images of the Radio Sky at Twenty-cm \citep[FIRST;][]{Becker95}
and NRAO VLA Sky Survey \citep[NVSS;][]{Condon98} 1.4-GHz surveys;
however the relatively shallow sensitivity levels of these
observations (0.75 and 2.25~mJy respectively) means that that these
surveys are dominated by classical radio galaxies, and contain few
star-forming galaxies.  We have been carrying out a series of deeper
radio surveys of the northern SWIRE fields, in order to extend the
multi-wavelength coverage currently available on star-forming galaxies
into radio frequencies.

In Section~\ref{sec:radioobservations} we present new observations of
the ELAIS-N2 region, taken at 610~MHz with the Giant Metrewave Radio
Telescope \citep[GMRT;][]{Ananthakrishnan05}.  We combine these
observations with our previous surveys of the ELAIS-N1 and Lockman
Hole regions \citep{Garn08EN1,Garn08LH} in
Section~\ref{sec:dataselection}, and create a sample of 510
star-forming galaxies with photometric redshifts that are detected in
the infrared and radio.  In Section~\ref{sec:results} we look at the
infrared / radio flux density ratio $q'_{\rm IR}$, and use deviations
away from the typical values seen for star-forming galaxies to
identify sources that may contain a significant excess of radio
emission due to the presence of an active galactic nucleus (AGN).  We
compare our sample to the local relationship between radio luminosity
and SFR given by \citet{Bell03}, and find that the majority of sources
are well described by this relationship, but that there are a small
fraction of sources where the SFR would be significantly
under-estimated by using the radio luminosity.  In
Section~\ref{sec:discussion} we test the assumption that the local
relationship between SFR and radio luminosity can be used to calculate
the SFRD of the Universe from galaxies at high redshift ($z=2$), and
from galaxies undergoing much more vigorous starbursts than are seen
in the local Universe.

Throughout this work a flat cosmology with the best-fitting parameters
from the Wilkinson Microwave Anisotropy Probe (WMAP) five-year data of
$\Omega_{\Lambda} = 0.74$ and H$_{0}$ = 72~km~s$^{-1}$~Mpc$^{-1}$ is
assumed \citep{Dunkley09}.

%%%%%%%%%%%%%%%%%%%%%%%%%%%%%%%%%%%%%%%%%%%%%%%%%%
\section{Radio observations}
\label{sec:radioobservations}
\subsection{The ELAIS-N2 region}
The ELAIS-N2 region was observed in 2006 July 15 and 17 with the GMRT.
Observations were made in two 16-MHz sidebands centred on 610~MHz,
each split into 128 spectral channels, with a 16.9~s integration time.
Thirteen pointings were observed, centred on 16$^{\rm h}$36$^{\rm
m}$48$^{\rm s}$, $+41\degr01'45''$ (J2000 epoch) and spaced by
36~arcmin in a hexagonal pattern, in a series of short interleaved
scans in order to maximise the $uv$ coverage.  The typical time spent
on each pointing was 22~min.  A nearby phase calibrator, J1613$+$342,
was observed for 4~min between every two or three target scans to
monitor any time-dependent phase and amplitude fluctuations of the
telescope.  Flux density calibration was performed using observations
of 3C48 or 3C286, at the beginning and end of each observing session.
The task {\sc setjy} was used to calculate 610-MHz flux densities of
29.4 and 21.1~Jy respectively, using the Astronomical Image Processing
Software ({\sc aips}) implementation of the \citet{Baars77} flux
density scale.

Data reduction was performed in a similar way to our previous GMRT
surveys \citep{Garn07,Garn08EN1,Garn08LH}.  Initial editing of the
data took place separately on each sideband with standard {\sc aips}
tasks, to remove bad baselines, antennas, and channels which were
suffering from large amounts of narrow band interference, along with
the first and last integration periods of each scan.  The flux
calibrators were used to create a bandpass correction for each
antenna.  In order to create a pseudo-continuum channel, five central
frequency channels were combined together, and an antenna-based phase
and amplitude calibration created using observations of J1613$+$342.
This calibration was applied to the original data, which was then
compressed into 11 channels of bandwidth 1.25~MHz, small enough that
bandwidth smearing was not a problem for our images.  The first and
last few spectral channels, which tended to be the noisiest, were
omitted from the data, leading to an total effective bandwidth of
13.75~MHz in each sideband.  Further flagging was performed on the
11-channel data set, and the two sidebands combined using {\sc uvflp}
\citep{Garn07} to improve the $uv$ coverage.  Baselines shorter than
1~k$\lambda$ were omitted from the imaging, since the GMRT has a large
number of small baselines which would otherwise dominate the beam
shape.

Each pointing was divided into 31 smaller facets, arranged in a
hexagonal grid, which were imaged individually with a separate assumed
phase centre.  The large total area imaged (a diameter of $1\fdg8$,
compared with the full width at half-maximum of the GMRT, which is
$\sim$0$\fdg74$) allows bright sources well away from the observed
phase centre to be cleaned from the images, while the faceting
procedure avoids phase errors being introduced due to the non-planar
nature of the sky.  All images were made with an elliptical restoring
beam of size $6.5 \times 5$~arcsec$^{2}$, PA $+70\degr$, with a pixel
size of 1.5~arcsec to ensure that the beam was oversampled.  The
images went through three iterations of phase self-calibration at 10,
3 and 1 min intervals, and then a final round of self-calibration
correcting both phase and amplitude errors, at 10~min intervals, with
the overall amplitude gain held constant in order not to alter the
flux density of sources.

The offset in the GMRT primary beam seen in our previous surveys was
again detected and corrected, using the method described in
\citet{Garn07}.  The 13 pointings were mosaicked together, taking into
account the offset primary beam and weighting the final mosaic
appropriately by the relative noise of each pointing.  The mosaic was
cut off at the point where the primary beam correction dropped to
20~per~cent of its central value, a radius of 0\fdg53 from the centre
of the outer pointings.

\begin{figure}
  \begin{center}
    \includegraphics[width=0.45\textwidth]{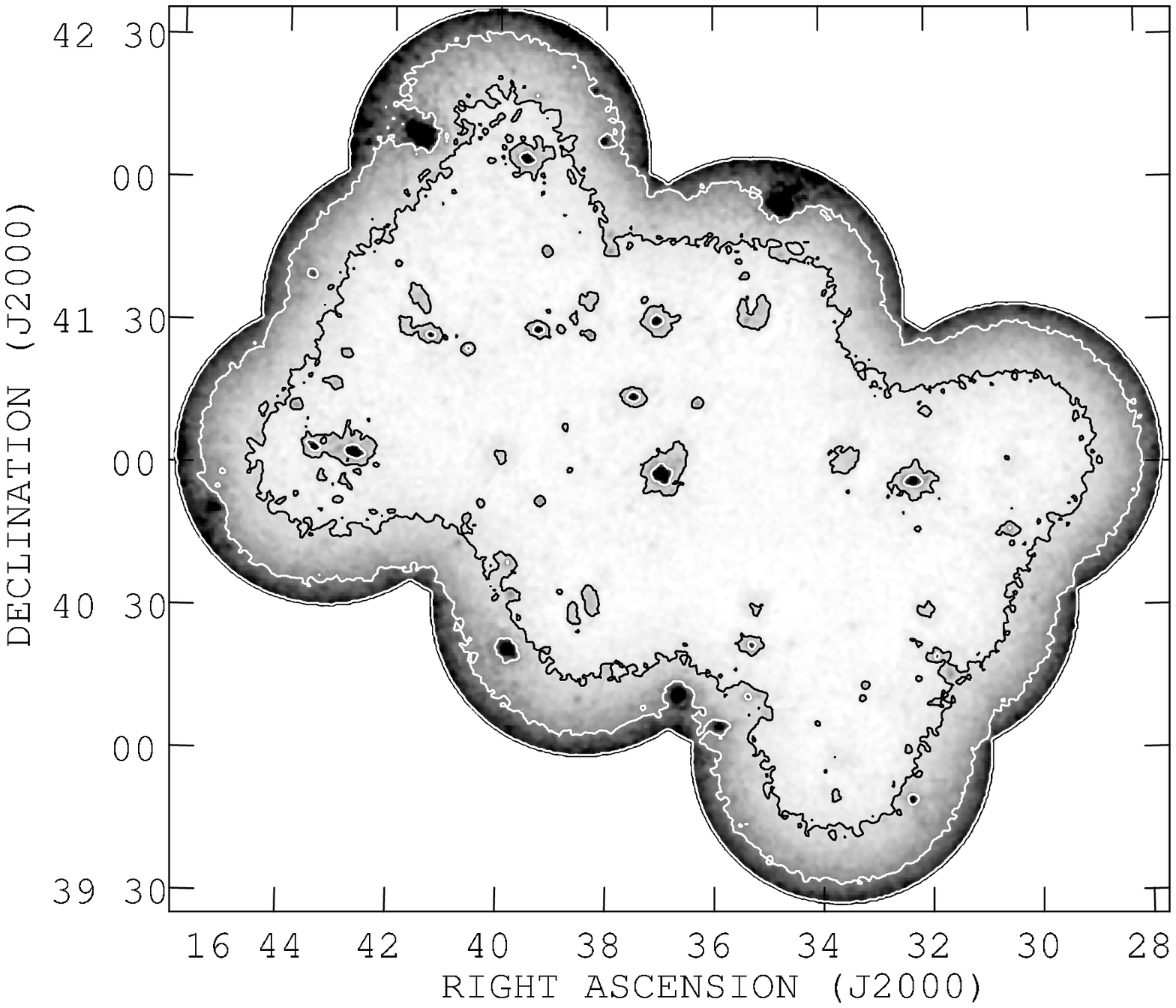}
    \caption{The rms noise map for the GMRT ELAIS-N2 survey.  The
    grey-scale ranges between 80 and 400~$\mu$Jy~beam$^{-1}$, and the
    contour levels are drawn at 120 and 240~$\mu$Jy~beam$^{-1}$ (black
    and white lines respectively).}
    \label{fig:EN2rms}
  \end{center}
\end{figure}

The rms noise for the individual pointings, before correction for the
GMRT primary beam, was between 83 and 90~$\mu$Jy~beam$^{-1}$, compared
with the expected noise level of 80~$\mu$Jy~beam$^{-1}$
\citep[calculated using Equation~1 from][]{Garn07}, and the rms noise
map for the mosaic is shown in Fig.~\ref{fig:EN2rms}.

%%%%%%%%%%%%%%%%%%%%%%%%%%%%%%%%%%%%%%%%%%%%%%%%%%
\subsection{Observations of ELAIS-N1 and the Lockman Hole}
610-MHz GMRT observations of the ELAIS-N1 field \citep{Garn08EN1} and
the central region of the Lockman Hole field \citep{Garn08LH} have
previously been presented.  A brief summary of the observations is
given here; for further details, the relevant papers should be
consulted.

The ELAIS-N1 survey covers $\sim$9~deg$^{2}$ (the majority of the {\it
Spitzer} observation region) with a resolution of
$6\times5$~arcsec$^{2}$, PA $+45^{\circ}$.  The rms noise level of the
majority of the survey was $\sim$70~$\mu$Jy~beam$^{-1}$ before primary
beam correction, with a smaller `deep' region having
$\sim$40~$\mu$Jy~beam$^{-1}$ rms.  

The Lockman Hole survey covers $\sim$5~deg$^{2}$, within the central
area of the region observed by {\it Spitzer}.  The survey resolution
was $6\times5$~arcsec$^{2}$, PA $+45^{\circ}$, and the noise level
$\sim$60~$\mu$Jy~beam$^{-1}$ over the majority of the region, with a
small area having significantly greater noise due to the presence of a
nearby bright radio source.

%%%%%%%%%%%%%%%%%%%%%%%%%%%%%%%%%%%%%%%%%%%%%%%%%%
\section{Data selection}
\label{sec:dataselection}
\subsection{Creating an infrared sample}
\label{sec:irsample}
The infrared data comes from the band-merged SWIRE source catalogues
of \citet{Surace05}, and the photometric redshift catalogue of
\citet{RowanRobinson08}.  In order to be included in the band-merged
catalogue, sources were required to be detected above a
signal-to-noise ratio (SNR) of 10 and 5 at 3.6~$\mu$m and 4.5~$\mu$m
(approximately equivalent to 10~$\mu$Jy in both bands).  There are
fewer sources present in the photometric redshift catalogue than the
band-merged catalogue, due to the need for sufficient spectral
information that a redshift could be determined.  Sources were
required to be present in both the band-merged and photometric
redshift catalogues, due to the need for a redshift, and the increased
amount of information present in the band-merged catalogue.  This
provided information on infrared flux densities in the four IRAC
bands, along with 24-$\mu$m and 70-$\mu$m flux densities, and
five-band optical photometry ($Ug'r'i'Z'$) from the Wide Field Survey
\citep[WFS;][]{McMahon01,Irwin01} for the ELAIS-N1 and ELAIS-N2
fields, with four-band ($U g'r'i'$) photometry for the Lockman Hole
from the SWIRE photometry program.

\begin{table*}
\begin{center}
\caption{The number of sources in each sub-sample -- see
  Sections~\ref{sec:irsample} and \ref{sec:radsample} for further
details on sample selection.}
\label{tab:samplesize}
\begin{tabular}{lcccc}
\hline
                         & ELAIS-N1 & ELAIS-N2 & Lockman Hole & Total\\
\hline
Infrared band-merged catalogue \citep{Surace05} & 282711 & 126056 & 323044 & 731811\\
Infrared photometric redshift catalogue \citep{RowanRobinson08} & 218117 & 125364 & 229238 & 572719\\
Band-merged and photometric catalogues & 121422 &  61610 & 132736 & 315768\\
Star-forming sources only, with SFR estimate &  49160 &  24831 &  42517 & 116508\\
Star-forming, and 24-$\mu$m detection &  11371 &   5725 &  12458 &  29554\\
Star-forming, 24-$\mu$m and 70-$\mu$m detections &   2470 &   1344 &   2939 &   6753\\
Star-forming, 24-$\mu$m and 70-$\mu$m detections, within radio coverage area &   2204 &   1317 &   1891 &   5412\\
\hline
Star-forming, 24-$\mu$m and 70-$\mu$m detections, plus 610-MHz radio
                         detection &    279 &     63 &    168 &    510\\
\hline
\end{tabular}
\end{center}
\end{table*}

Sources must also be classified as `star-forming', and have an
estimate of the SFR.  This classification and estimate comes from the
template fitting performed by \citet{RowanRobinson08} during their
photometric redshift calculations, and rejects some sources that are
identified as having their infrared energetics dominated by AGN
activity.  An infrared detection at both 24 and 70~$\mu$m is also
required in order to constrain the thermal emission from dust, and to
have sufficient information on the region of the infrared SED where
the bulk of the bolometric energy output of star-forming galaxies is
seen.  Selecting galaxies at both 24 and 70~$\mu$m increases the
accuracy of the infrared bolometric luminosity calculation performed
by \citet{RowanRobinson08}, which would otherwise have a factor of
$\sim$2 uncertainty, and improves the accuracy of their SFR estimates.
Finally, we require the sources to lie within the area that was
observed in one of the GMRT radio surveys.  The numbers of sources in
each field that satisfy the selection criteria are given in
Table~\ref{tab:samplesize}.

%%%%%%%%%%%%%%%%%%%%%%%%%%%%%%%%%%%%%%%%%%%%%%%%%%
\subsection{Matching to the radio data}
\label{sec:radsample}
In creating the public catalogues for the ELAIS-N1 and Lockman Hole
GMRT surveys, a two-stage selection criterion with an increased SNR
requirement near to bright sources was implemented \citep[for more
details see][]{Garn08EN1} -- this was due to problems encountered from
residual phase errors near to the brightest radio sources in the
survey.  For this study, we have known source positions from the
infrared and optical detections, and so relax this criterion to a SNR
of 4.  Radio sources were matched to the infrared data if the source
centres were within a distance $r$ of each other -- if there were
potential multiple matches, the closest match was selected.  There
will be some incorrect associations between the infrared and radio
catalogues -- we estimated this value by shifting the coordinates of
all radio sources by 10~arcmin in declination and seeing how many
(incorrect) associations were then made at the same matching radius
$r$.  We required the false matching rate to be less than 5~per~cent,
which corresponded to $r$ = 2.5~arcsec.  The number of
galaxies with a radio detection within 2.5~arcsec are listed in
Table~\ref{tab:samplesize}, with a total of 510 galaxies in all three
survey fields that satisfied the infrared and radio selection
criteria.

%%%%%%%%%%%%%%%%%%%%%%%%%%%%%%%%%%%%%%%%%%%%%%%%%%
\subsection{Sample selection biases}
\label{sec:selectionbiases}
We have rejected sources from the infrared-selected samples that were
classified by \citet{RowanRobinson08} as having optical and infrared
photometry that is best fitted by an AGN template; however there are
several examples of galaxies that undergo concurrent AGN and star
formation activity \citep*[e.g.][]{Richards07,Norris07a} which will
not necessarily be identified during this process, and this selection
will not remove sources which have a significant contribution to their
radio emission from AGN activity, but are not clearly AGN-like at
shorter wavelengths.  We return to the question of AGN contaminants in
Section~\ref{sec:resultsq}.

Selecting galaxies at both 24 and 70~$\mu$m places a significant
restriction on the sample size, as seen in Table~\ref{tab:samplesize}.
Approximately 25~per~cent of the star-forming galaxies with an
estimate of SFR were detected in the 24-$\mu$m band, and of these
galaxies, approximately 25~per~cent were detected at 70~$\mu$m.  The
infrared emission at 24~$\mu$m is dominated by transiently heated
polycyclic aromatic hydrocarbon (PAH) molecules \citep[e.g.][and
references therein]{Leger84,Puget89,Draine03} and is strongly related
to the amount of ongoing star formation, while the emission at
70~$\mu$m is dominated by thermal emission from large dust grains.
Both detections are considered necessary in order to be sure of the
accuracy of the template fitting and SFR estimation.

\begin{figure}
  \centerline{\subfigure[]{
    \includegraphics[width=0.45\textwidth]{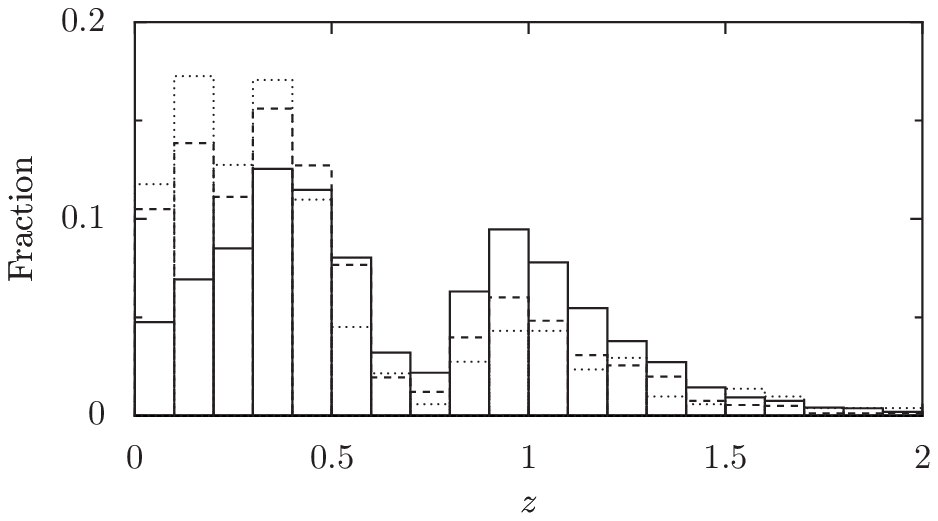}}}
  \centerline{\subfigure[]{
    \includegraphics[width=0.45\textwidth]{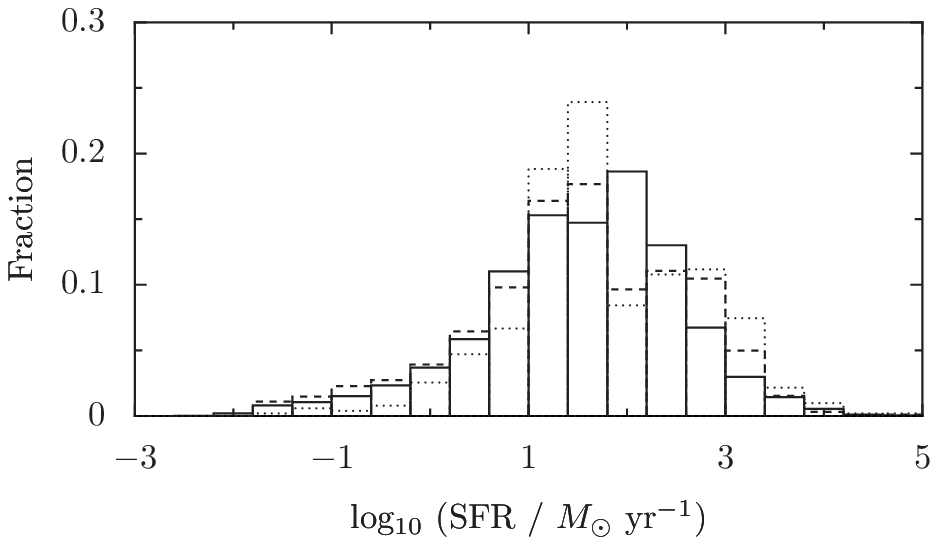}}}
  \caption{The fractional distribution of (a) redshift and (b) SFR for
    galaxies selected at 24~$\mu$m (solid line), 24 and 70~$\mu$m
    (dashed line), and 24~$\mu$m, 70~$\mu$m and 610~MHz (dotted line)
    -- see Section~\ref{sec:selectionbiases} for further details.}
  \label{fig:distributions}
\end{figure}

\begin{figure}
  \begin{center}
    \includegraphics[width=0.45\textwidth]{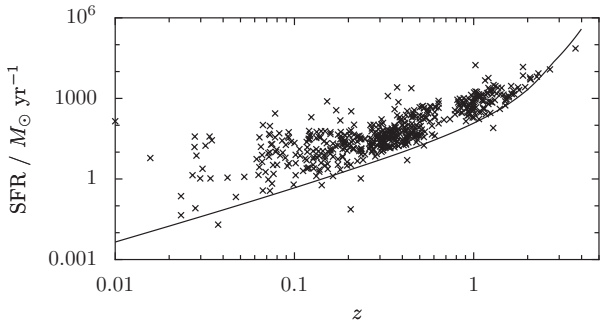}
    \caption{The relationship between $z$ and SFR for the 510 galaxies
      detected at 24~$\mu$m, 70~$\mu$m and 610~MHz.  The solid line
      represents the limiting SFR detectable at different redshifts,
      using the relationship between monochromatic 70-$\mu$m flux
      density and SFR given by \citet{Rieke09}.}
    \label{fig:sfrz}
  \end{center}
\end{figure}

Fig.~\ref{fig:distributions}a shows the redshift distribution of
24-$\mu$m selected galaxies.  The bulk of galaxies are at $z < 0.5$,
although there is a secondary peak at $z\sim1$ due to the shifting of
the 11.7-$\mu$m PAH feature into the 24-$\mu$m band
\citep{RowanRobinson08}.  The 70-$\mu$m flux limit for the SWIRE
surveys is significantly higher than the 24-$\mu$m limit \citep[30~mJy
compared with 0.45~mJy;][]{Surace05}, and by selecting galaxies with
70-$\mu$m fluxes, the sample is biased towards galaxies that are
brighter at mid-infrared wavelengths, and have high SFR.  The
secondary redshift peak at $z\sim1$ is reduced in strength for the
sample selected at 24-$\mu$m and 70-$\mu$m, but remains prominent --
this is due to the global decrease in SFRD of the Universe, seen
between $z\sim2$ and the current day
\citep[e.g.][]{Lilly96,Madau96,Hopkins06}, a consequence of which is
that the highest SFR galaxies are likely to be found at higher
redshift.  The SFR distribution of the two samples is given in
Fig.~\ref{fig:distributions}b, and shows that a greater fraction of
galaxies in the 24-$\mu$m and 70-$\mu$m selected sample have SFR $\geq
100$~$M_{\odot}$~yr$^{-1}$ than when only a 24-$\mu$m detection is
required.  Fig.~\ref{fig:sfrz} illustrates that high-SFR sources are
more likely to be found at higher $z$ in our sample.  The upper left
of the diagram is relatively unpopulated due to the lower space
density of high SFR sources in the local Universe compared to at
$z\sim2$, while the lower right of the diagram is unpopulated
principally because of the flux limit of the 70-$\mu$m data used in
the \citet{RowanRobinson08} catalogue.  The curved line indicates the
minimum detectable SFR at different redshifts, based upon the
70-$\mu$m flux only, using the relationship between a monochromatic
MIPS flux density and SFR given by \citet{Rieke09}.  There are a few
sources seen below this limit, due to the method used by
\citet{RowanRobinson08} to estimate SFRs (see
Section~\ref{sec:overestsfr}), which considers more information than
just a single 70-$\mu$m detection.  Note that the shape of the SFR
limit depends both on the intrinsic limiting luminosity for a
flux-limited sample, and the $k$-correction caused by the slope of the
mid-infrared SED for a typical starburst galaxy.

The photometric redshift catalogue of \citet{RowanRobinson08} contains
a considerable number of sources with an inferred SFR of
$>1000$~$M_{\odot}$~yr$^{-1}$.  Such extreme SFRs are unusual, and for
many sources are likely to be unphysical, but are not unknown
\citep[see, e.g.][for a discussion of a spectroscopically-confirmed
source with SFR $\sim4000$~$M_{\odot}$~yr$^{-1}$]{Capak08}.  We note
that SFRs derived from either 24-$\mu$m or 70-$\mu$m fluxes are
consistent with these high values for most of the extreme SFR sources
in our sample (see Section~\ref{sec:overestsfr}), although the
greatest SFR that \citet{Rieke09} model is
$\sim1600$~$M_{\odot}$~yr$^{-1}$, and inferred SFRs above this value
will be very poorly constrained.  We also note that SFRs derived
independently from radio data are consistent with these high values
(see Section~\ref{sec:resultssfr}), although in both cases an error in
the calculated photometric redshift may be leading to a significant
over-estimation of the SFR.  The small number of sources in our sample
with a very high inferred SFR do not affect our later conclusions on
variations in the relationship between radio luminosity and SFR with
redshift.

Selecting galaxies with photometric redshifts, rather than requiring
spectroscopic redshifts, improves the homogeneity of the sample -- all
galaxies with sufficient photometry will be included in the study,
rather than limiting the sample to galaxies that are sufficiently
bright at one wavelength that they were selected for spectroscopic
followup.  Photometric redshifts are inherently less accurate than
spectroscopic equivalents -- however, there were considerable numbers
of spectroscopic redshifts available for validation in the SWIRE
fields, and \citet{RowanRobinson08} estimate an typical error of
4~per~cent in $(z_{\rm phot} - z_{\rm spec})/(1 + z_{\rm spec})$, with
$\sim$1~per~cent of the photometric redshifts being catastrophically
incorrect.  Where both spectroscopic and photometric redshifts are
available, we use the spectroscopically determined value, and we
return to the possibility of catastrophic redshift errors in
Section~\ref{sec:overestsfr}.

The radio surveys used in this work are much less sensitive to
star-forming galaxies than the infrared surveys, as can be seen by the
low number of source detections in the bottom row of
Table~\ref{tab:samplesize}.  The percentage of infrared-selected
galaxies that are detected at 610~MHz is 4.8~per~cent (ELAIS-N2),
8.9~per~cent (Lockman Hole) and 12.7~per~cent (ELAIS-N1), with the
fields that have lower 610-MHz noise levels having a greater
percentage of source matches.  Due to the positive correlation that is
seen between radio luminosity and SFR \citep[e.g.][and see
Section~\ref{sec:resultssfr}]{Condon90,Bell03}, additionally requiring
a detection at 610~MHz will bias the sample against galaxies with very
low SFR.  This can be seen in Fig.~\ref{fig:distributions}b -- the
fraction of galaxies with SFR $<$ 1~$M_{\odot}$~yr$^{-1}$ is
significantly reduced compared with the infrared-only samples.  A
study of sources below the radio detection limit, requiring a
statistical process such as median source stacking
\citep[e.g.][]{Boyle07,Beswick08}, is beyond the scope of this paper,
but has been discussed in a separate work \citep{Garn09stacking}.  We
will return to the galaxies where a non-detection in the radio is
significant in Section~\ref{sec:radioquiet}.

%%%%%%%%%%%%%%%%%%%%%%%%%%%%%%%%%%%%%%%%%%%%%%%%%%
\section{Results}
\label{sec:results}
\subsection{Identifying potential AGN contaminants}
\label{sec:resultsq}
\begin{figure}
  \centerline{\subfigure[]{
  \includegraphics[width=0.45\textwidth]{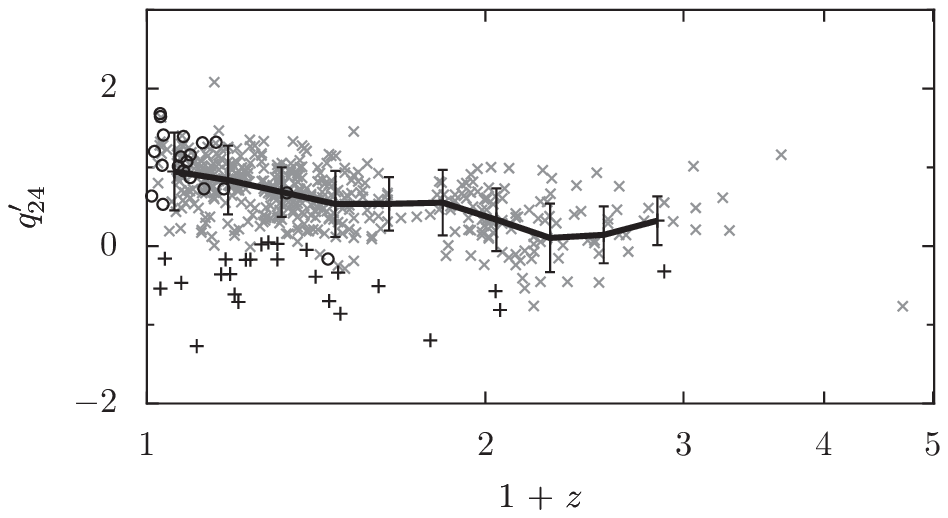}}}
  \centerline{\subfigure[]{
  \includegraphics[width=0.45\textwidth]{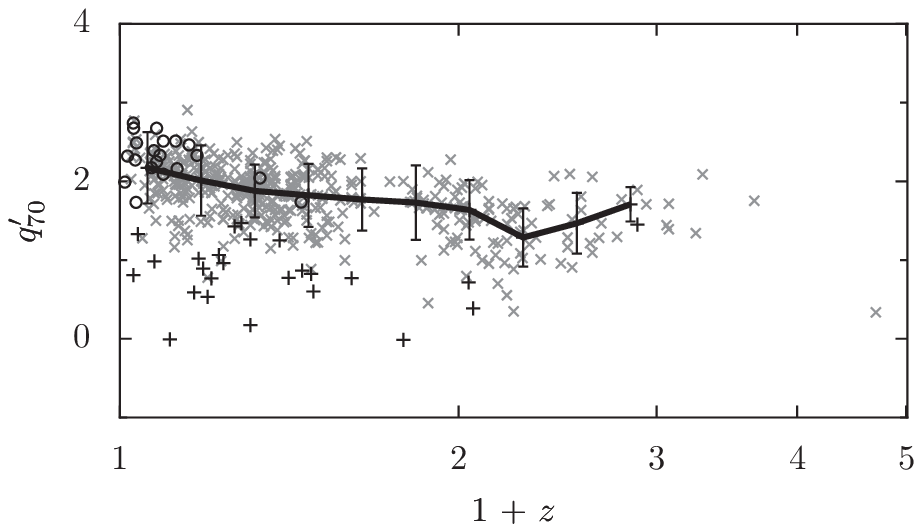}}}
  \caption{Variation in (a) $q'_{24}$ and (b) $q'_{70}$ with redshift,
  along with the local median value of $q'_{\rm IR}$ (black line with
  error bars).  Sources which have been identified as being AGN-like
  from their deviation below the median value of $q'_{24}$ are
  indicated by crosses, `+' -- see Section~\ref{sec:resultsq} for
  further details.  Open circles, `$\circ$', mark the sources that
  have been identified as being radio-quiet compared with their SFR --
  see Section~\ref{sec:radioquiet} for further details.}
  \label{fig:q}
\end{figure}

There is a tight relationship between the infrared and radio
luminosities of galaxies that are predominantly powered by
star-formation processes \citep[the `infrared / radio correlation';
e.g.][]{Helou85,Condon91,Murphy06}, which is not followed by sources
that have significant amounts of AGN activity \citep{Sopp91,Roy98}.
The correlation has previously been used as a discriminant between the
two types of source \citep[e.g.][]{Ibar08,Seymour08}, and in order to
use this technique on our sample, we consider the monochromatic
$q'_{\rm IR}$ parameter given by
\begin{equation}
  q'_{\rm IR} = {\rm log}_{10} \left(\frac{S_{\rm
  IR}}{S_{610}}\right),
\end{equation}
where $S_{\rm IR}$ is the infrared flux density in either the
24-$\mu$m or 70-$\mu$m bands, and $S_{610}$ is the radio flux density
at 610~MHz.  Note that this definition is slightly different from
previous works \citep[e.g.][]{Appleton04} which typically use 1.4-GHz
flux densities.  Fig.~\ref{fig:q} shows the variation in $q'_{24}$ and
$q'_{70}$ with redshift for the galaxies in this sample.  The
dispersion in the individual $q'_{\rm IR}$ values is greater than
typically seen at 1.4~GHz \citep{Appleton04}.  The principal reason
for this is that for galaxies which have {\it both} AGN and
star-forming activity, the 610-MHz flux density mainly stems from the
nuclear source, while the 1.4-GHz flux density comes mainly from star
formation activity \citep{Magliocchetti08}.  AGNs were identified by
finding sources that were a better fit to the selected AGN templates
of \citet{RowanRobinson08} than to elliptical galaxy templates, as
discussed in Section~\ref{sec:dataselection}; these were then removed
from the sample.  Any sources that have their radio luminosity
dominated by an AGN, but do not clearly show up as having infrared AGN
templates, will not have been removed through this method.  These
`missed' AGN sources will be radio bright, and therefore have a low
value of $q'_{\rm IR}$.

\begin{figure*}
 \centerline{
   \includegraphics[width=0.07\textwidth]{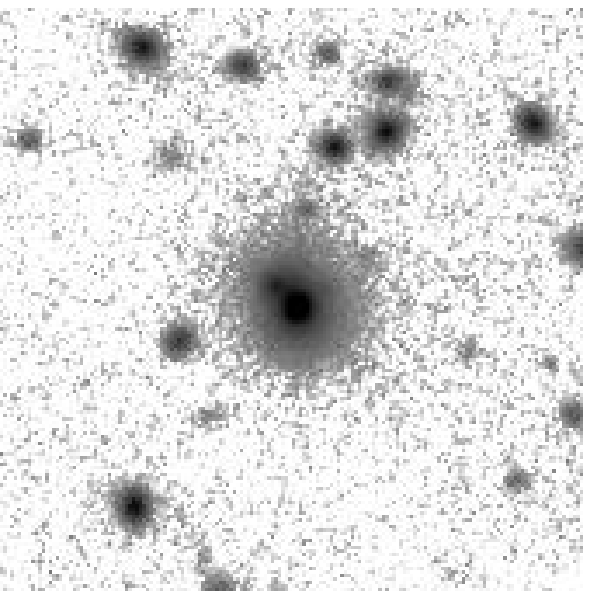}
   \includegraphics[width=0.07\textwidth]{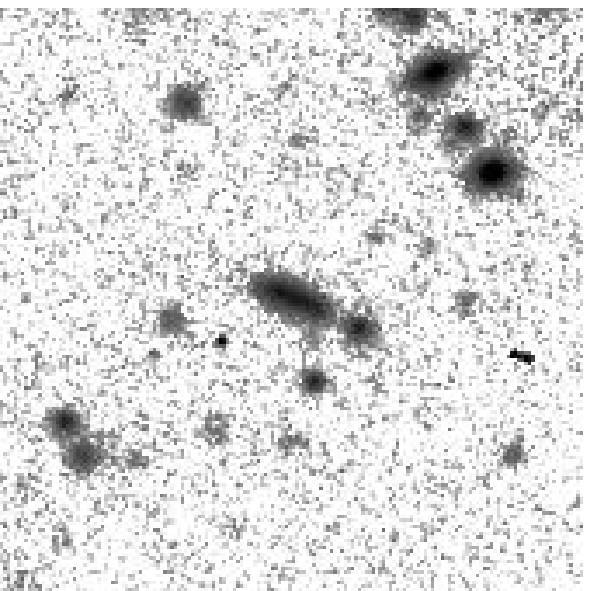}
   \includegraphics[width=0.07\textwidth]{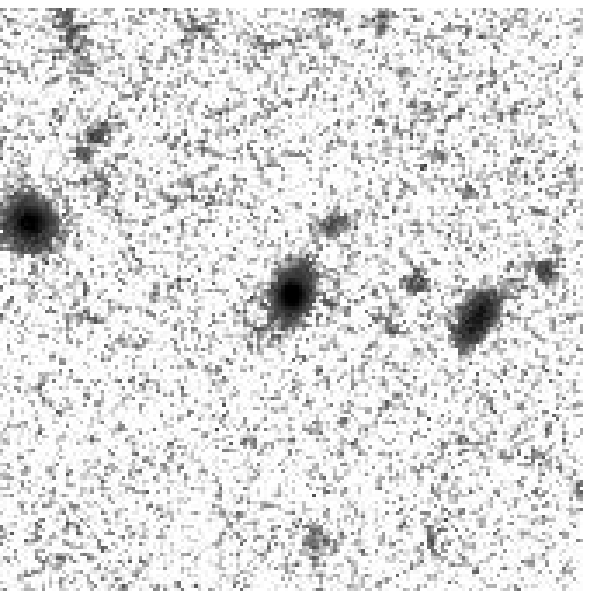}
   \includegraphics[width=0.07\textwidth]{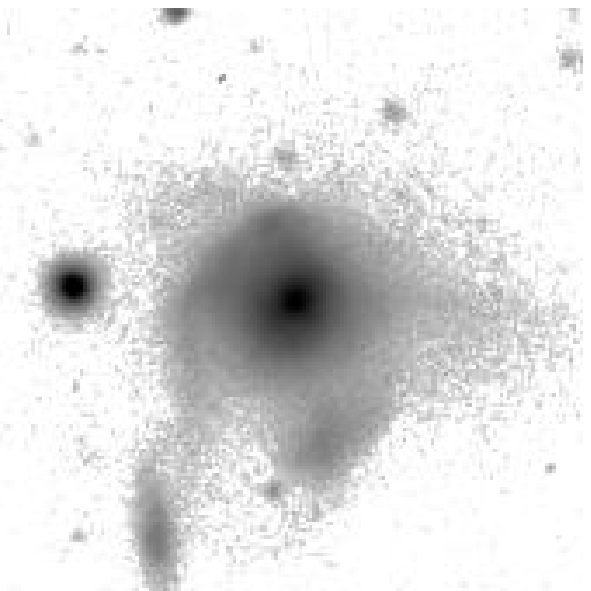}
   \includegraphics[width=0.07\textwidth]{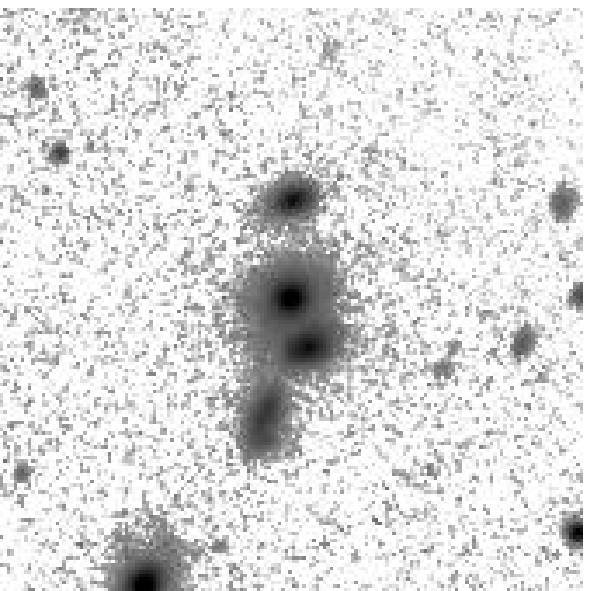}
   \includegraphics[width=0.07\textwidth]{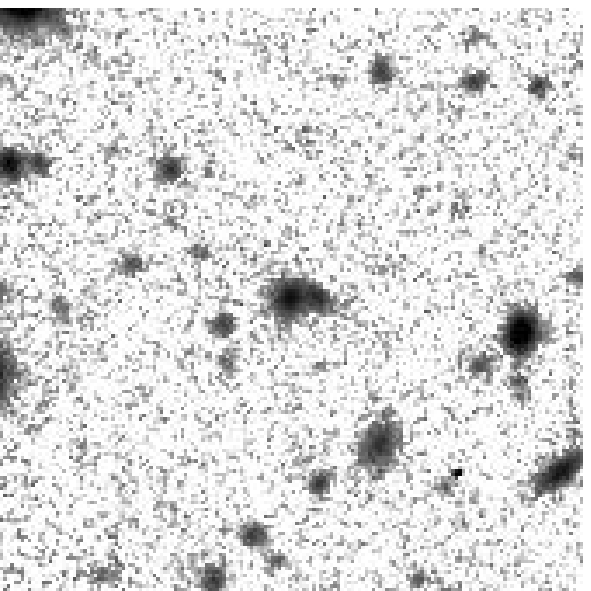}
   \includegraphics[width=0.07\textwidth]{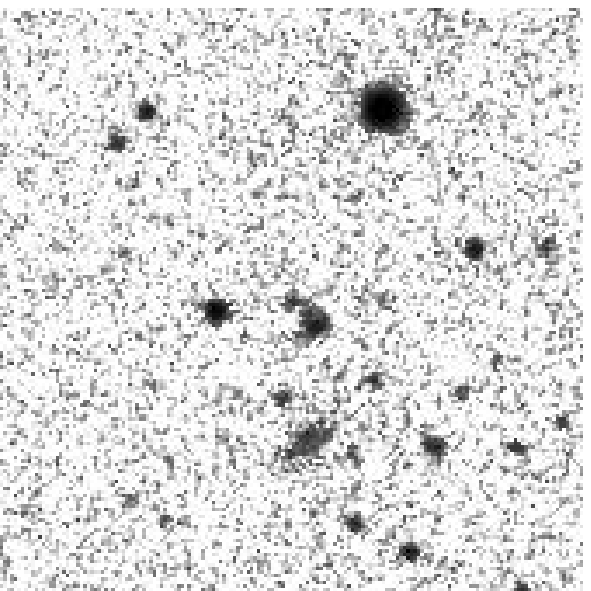}
   \includegraphics[width=0.07\textwidth]{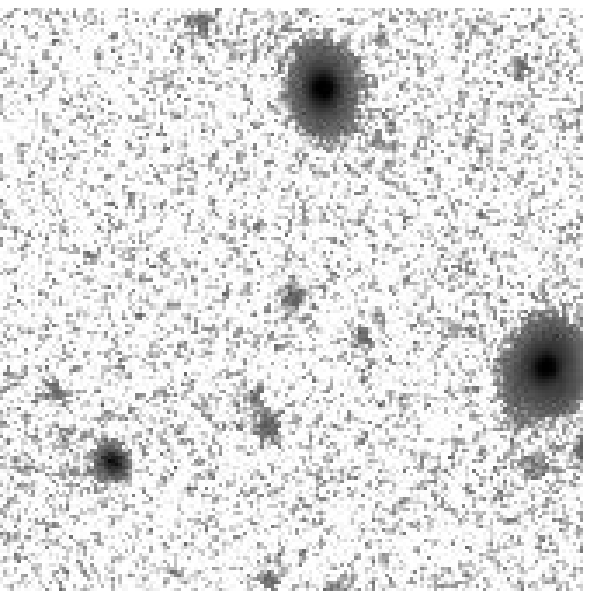}
   \includegraphics[width=0.07\textwidth]{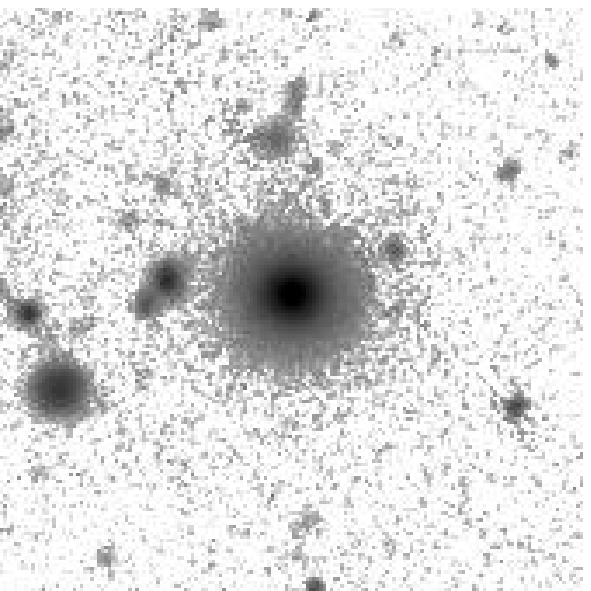}
   \includegraphics[width=0.07\textwidth]{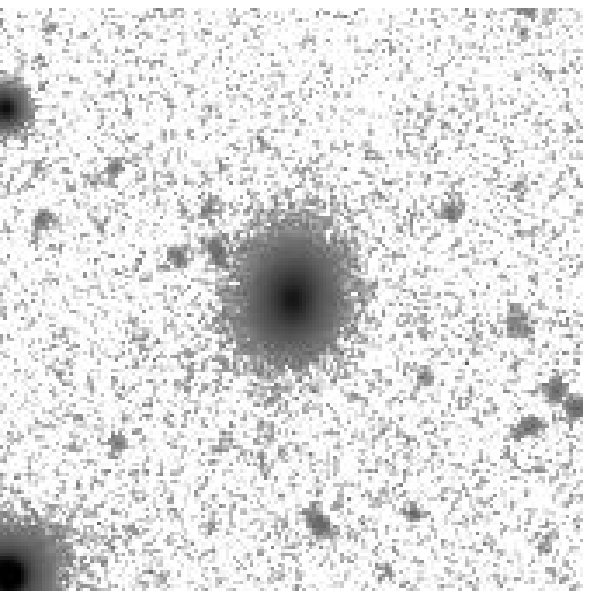}
   \includegraphics[width=0.07\textwidth]{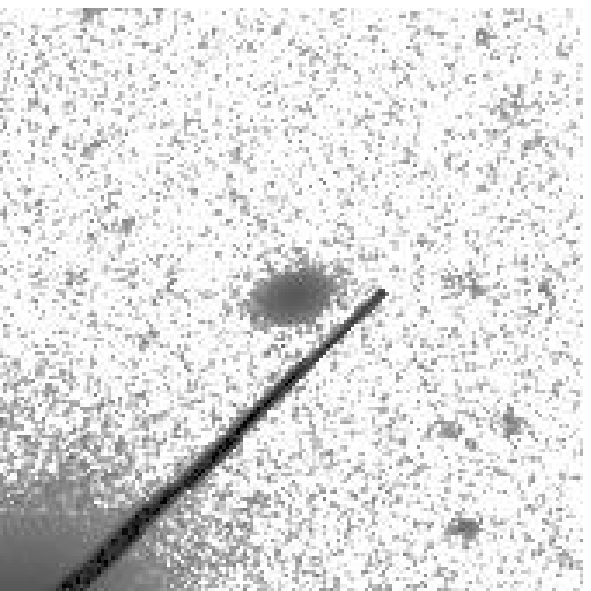}
   \includegraphics[width=0.07\textwidth]{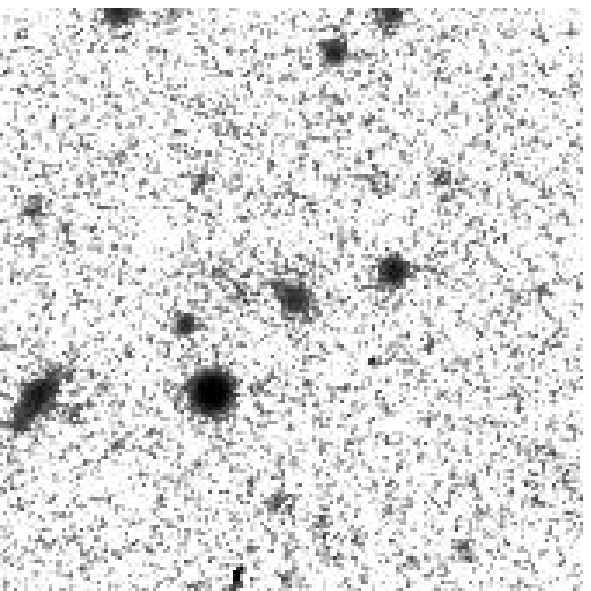}
   \includegraphics[width=0.07\textwidth]{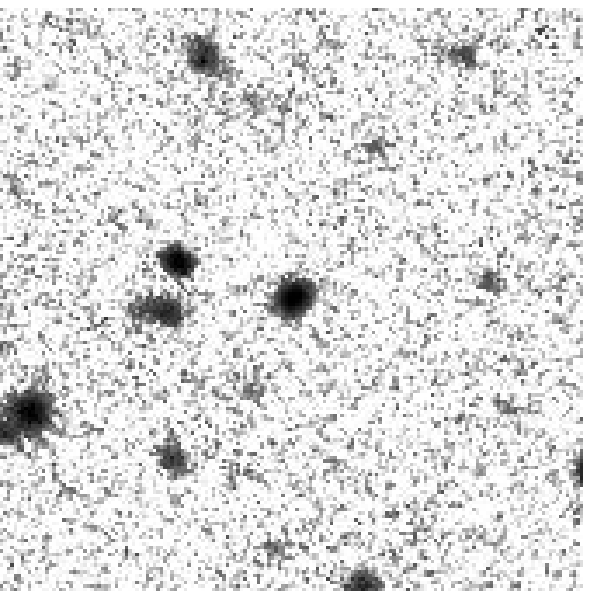}}
 \centerline{
   \includegraphics[width=0.07\textwidth]{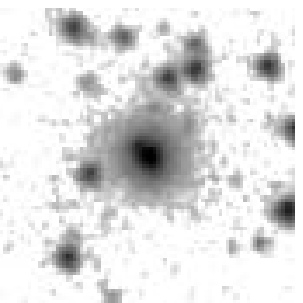}
   \includegraphics[width=0.07\textwidth]{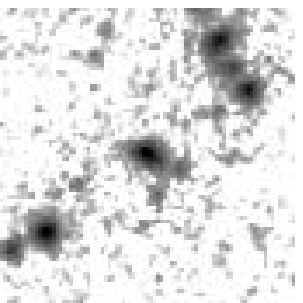}
   \includegraphics[width=0.07\textwidth]{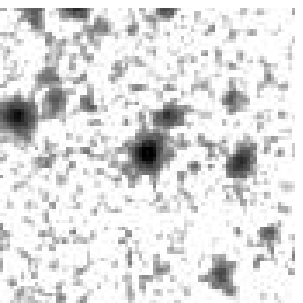}
   \includegraphics[width=0.07\textwidth]{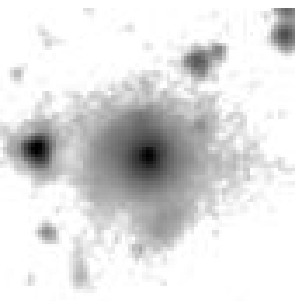}
   \includegraphics[width=0.07\textwidth]{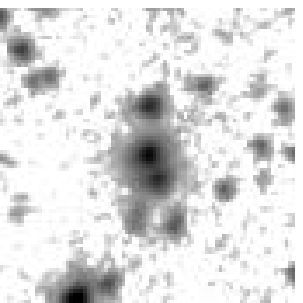}
   \includegraphics[width=0.07\textwidth]{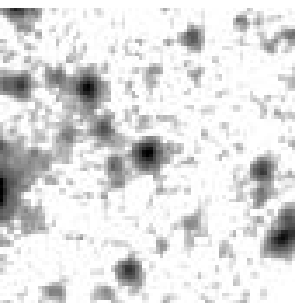}
   \includegraphics[width=0.07\textwidth]{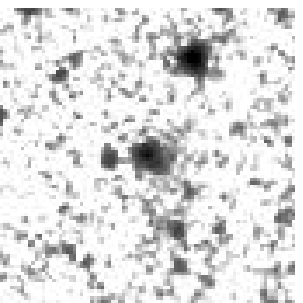}
   \includegraphics[width=0.07\textwidth]{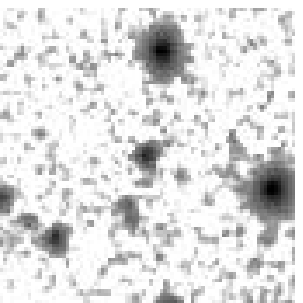}
   \includegraphics[width=0.07\textwidth]{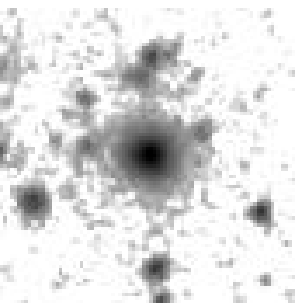}
   \includegraphics[width=0.07\textwidth]{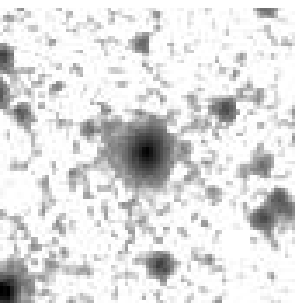}
   \includegraphics[width=0.07\textwidth]{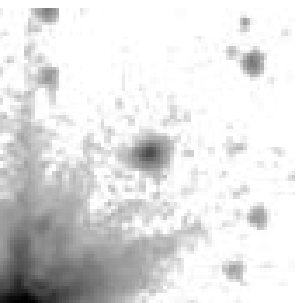}
   \includegraphics[width=0.07\textwidth]{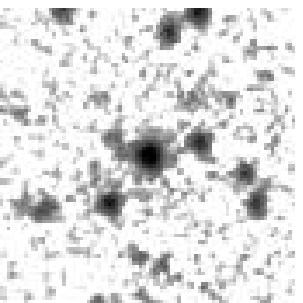}
   \includegraphics[width=0.07\textwidth]{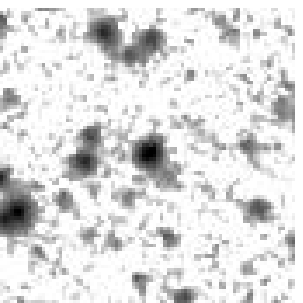}}
 \centerline{
   \includegraphics[width=0.07\textwidth]{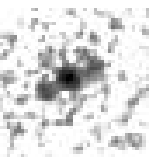}
   \includegraphics[width=0.07\textwidth]{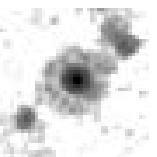}
   \includegraphics[width=0.07\textwidth]{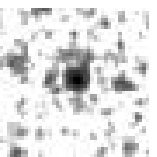}
   \includegraphics[width=0.07\textwidth]{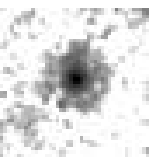}
   \includegraphics[width=0.07\textwidth]{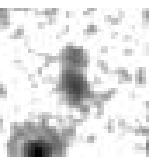}
   \includegraphics[width=0.07\textwidth]{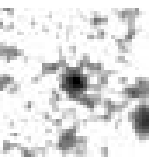}
   \includegraphics[width=0.07\textwidth]{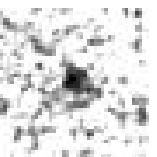}
   \includegraphics[width=0.07\textwidth]{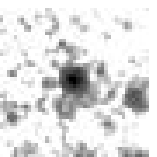}
   \includegraphics[width=0.07\textwidth]{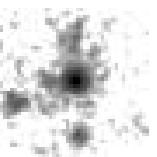}
   \includegraphics[width=0.07\textwidth]{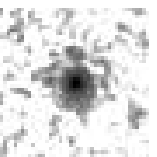}
   \includegraphics[width=0.07\textwidth]{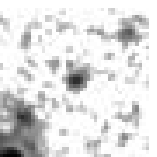}
   \includegraphics[width=0.07\textwidth]{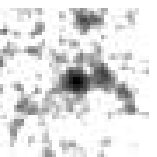}
   \includegraphics[width=0.07\textwidth]{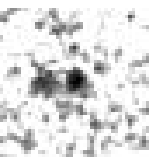}}
\vskip0.2cm
  \centerline{
   \includegraphics[width=0.07\textwidth]{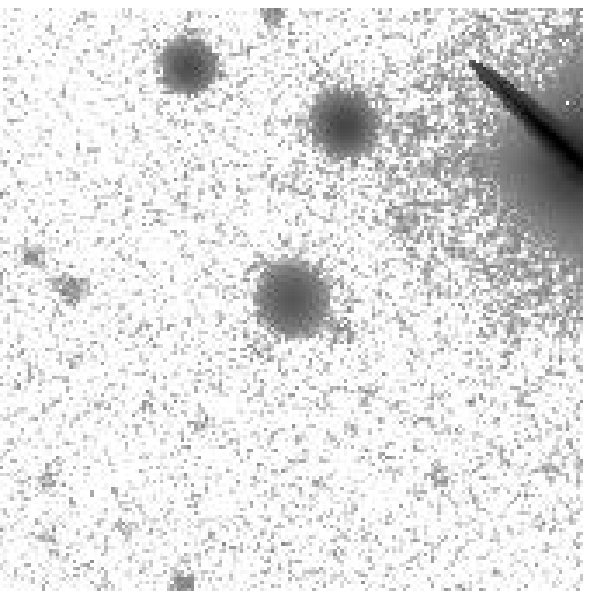}
   \includegraphics[width=0.07\textwidth]{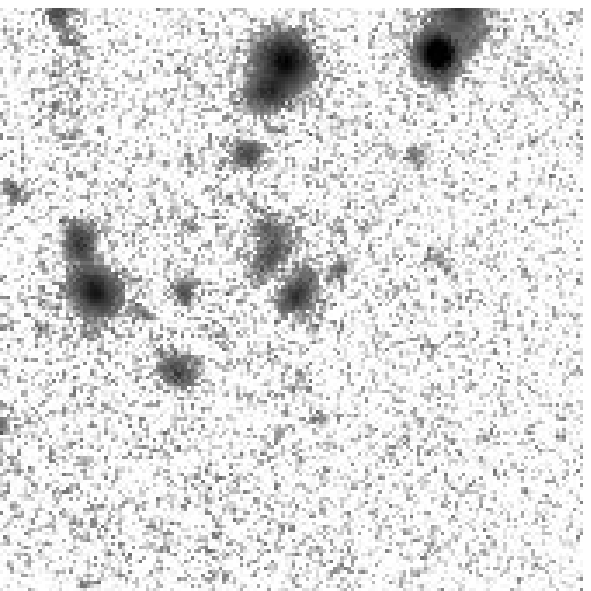}
   \includegraphics[width=0.07\textwidth]{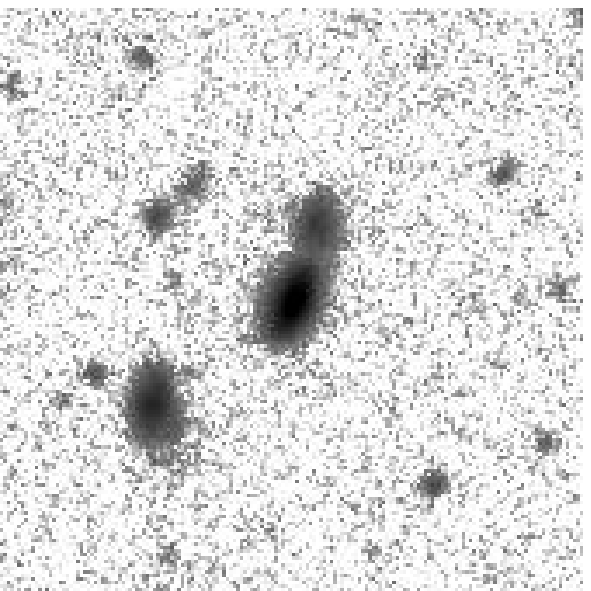}
   \includegraphics[width=0.07\textwidth]{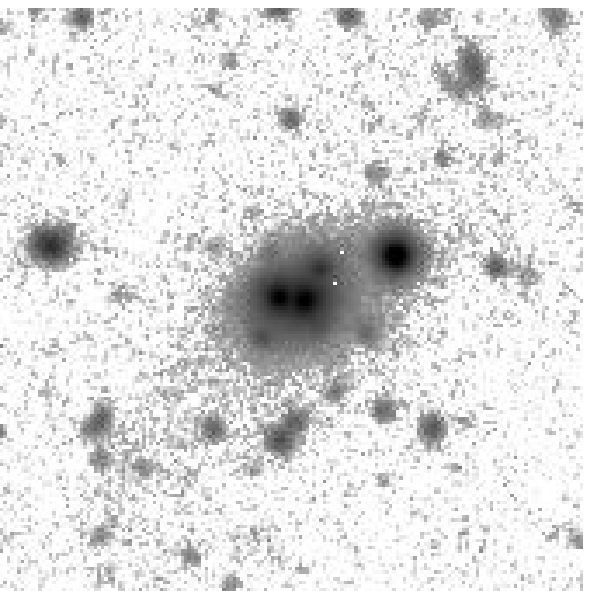}
   \includegraphics[width=0.07\textwidth]{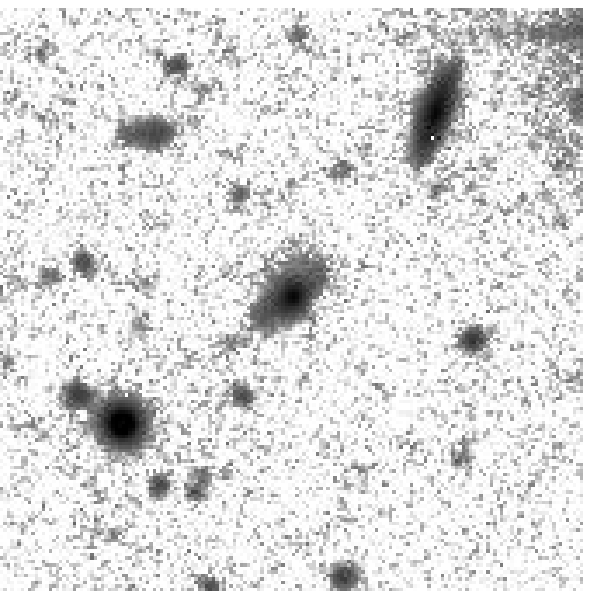}
   \includegraphics[width=0.07\textwidth]{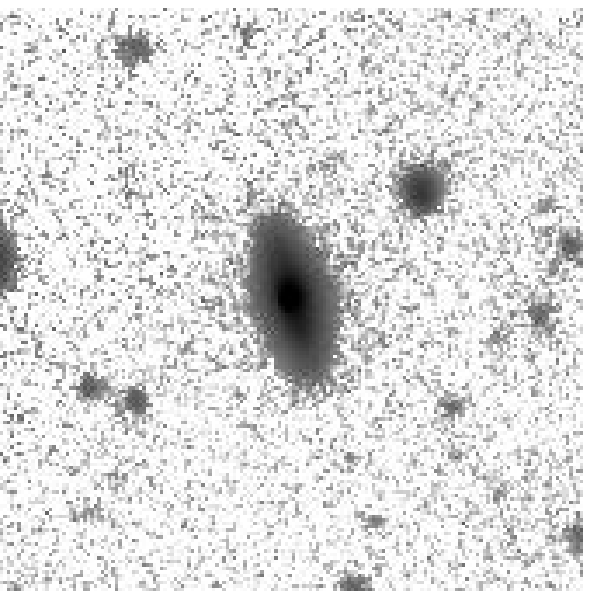}
   \includegraphics[width=0.07\textwidth]{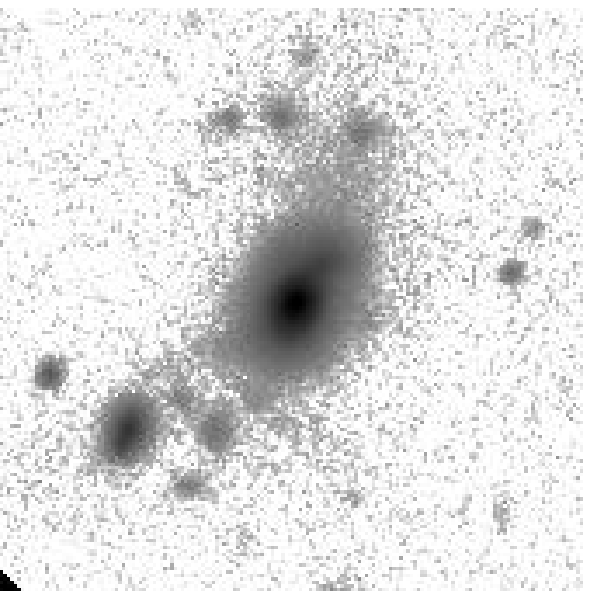}
   \includegraphics[width=0.07\textwidth]{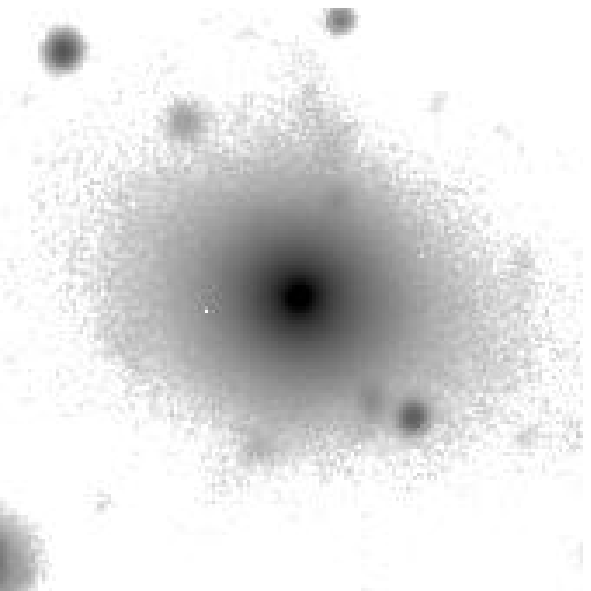}
   \includegraphics[width=0.07\textwidth]{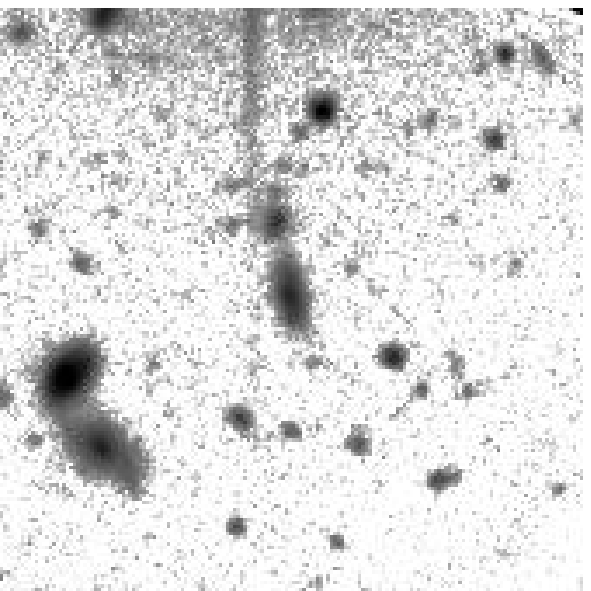}
   \includegraphics[width=0.07\textwidth]{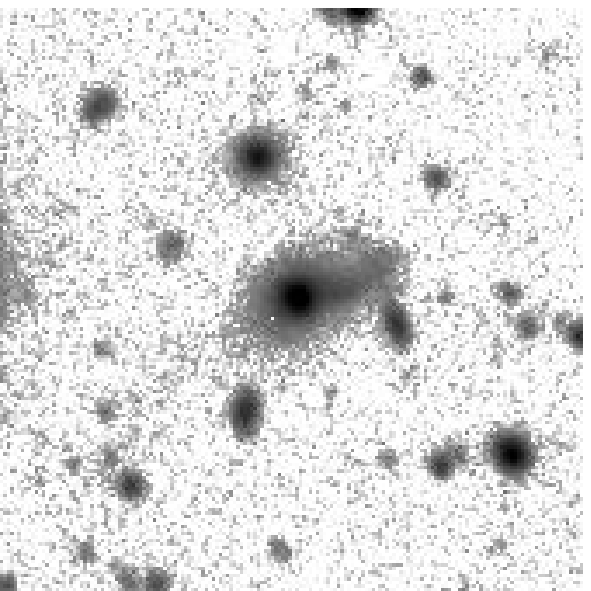}
   \includegraphics[width=0.07\textwidth]{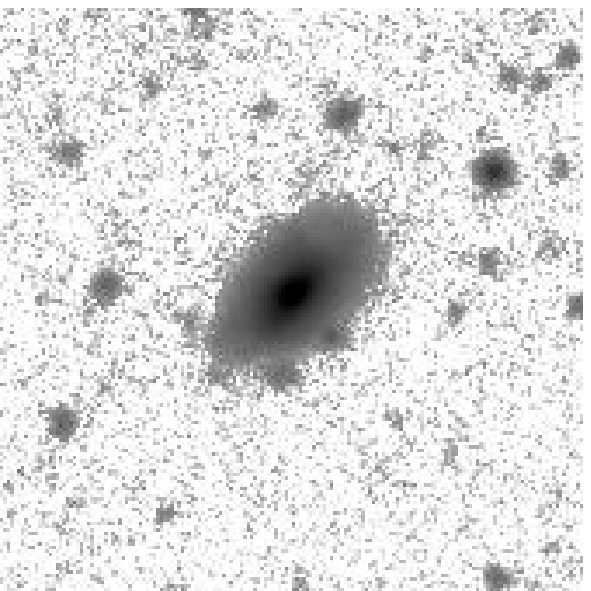}
   \includegraphics[width=0.07\textwidth]{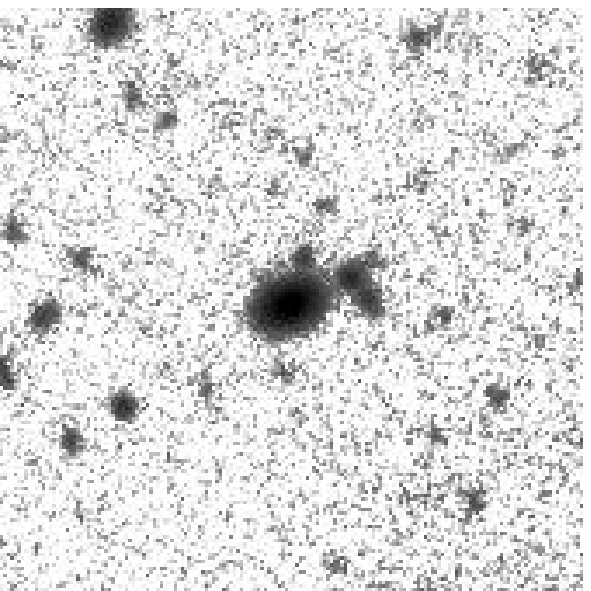}}
 \centerline{
   \includegraphics[width=0.07\textwidth]{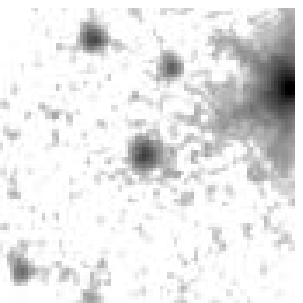}
   \includegraphics[width=0.07\textwidth]{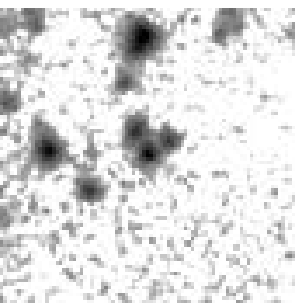}
   \includegraphics[width=0.07\textwidth]{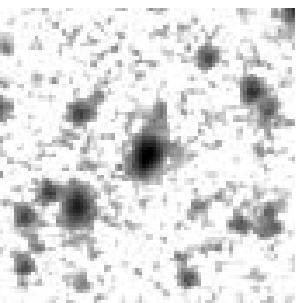}
   \includegraphics[width=0.07\textwidth]{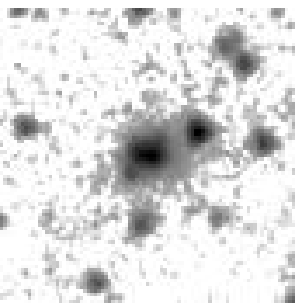}
   \includegraphics[width=0.07\textwidth]{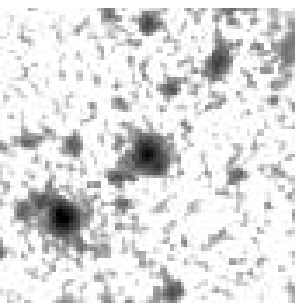}
   \includegraphics[width=0.07\textwidth]{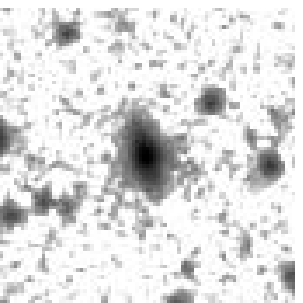}
   \includegraphics[width=0.07\textwidth]{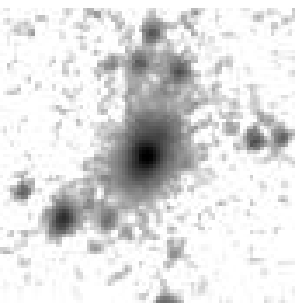}
   \includegraphics[width=0.07\textwidth]{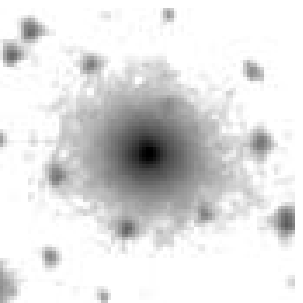}
   \includegraphics[width=0.07\textwidth]{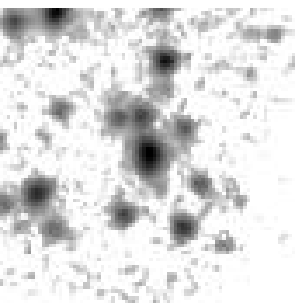}
   \includegraphics[width=0.07\textwidth]{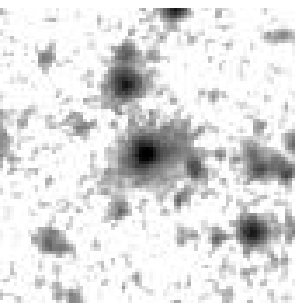}
   \includegraphics[width=0.07\textwidth]{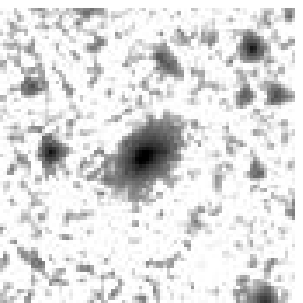}
   \includegraphics[width=0.07\textwidth]{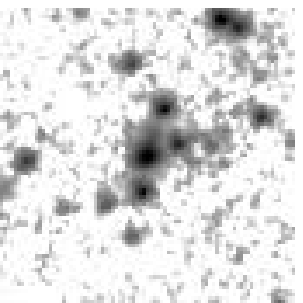}}
 \centerline{
   \includegraphics[width=0.07\textwidth]{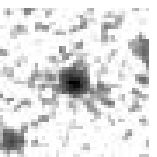}
   \includegraphics[width=0.07\textwidth]{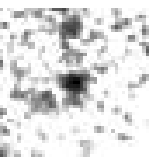}
   \includegraphics[width=0.07\textwidth]{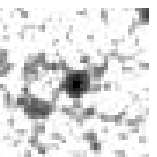}
   \includegraphics[width=0.07\textwidth]{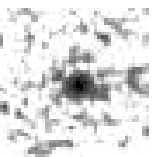}
   \includegraphics[width=0.07\textwidth]{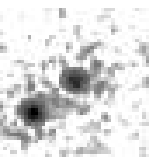}
   \includegraphics[width=0.07\textwidth]{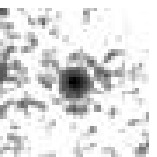}
   \includegraphics[width=0.07\textwidth]{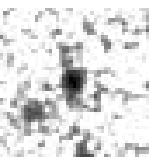}
   \includegraphics[width=0.07\textwidth]{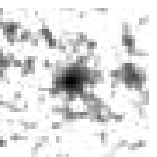}
   \includegraphics[width=0.07\textwidth]{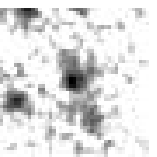}
   \includegraphics[width=0.07\textwidth]{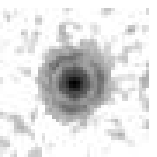}
   \includegraphics[width=0.07\textwidth]{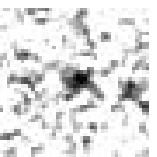}
   \includegraphics[width=0.07\textwidth]{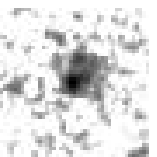}}
  \caption{Optical and infrared images of the 25 AGN-like sources,
  retrieved from the NASA/IPAC infrared science archive.  All images
  are $50\times50$~arcsec$^{2}$ in size.  Top row: $r'$-band images.
  Middle row: 3.6-$\mu$m images.  Bottom row: 24-$\mu$m images.}
  \label{fig:agn}
\end{figure*}

We have not converted the data in Fig.~\ref{fig:q} to a consistent
rest-frame wavelength (a `$k$-correction'), and so the observed value
of $q'_{\rm IR}$ will vary with the redshift of the galaxy, as
different portions of the redshifted SED enter into our observation
bands.  In order to quantify how the typical value of $q'_{\rm IR}$
varies, we calculate the median value of $q'_{\rm IR}$ in ten
logarithmically-spaced redshift bins between $z$ = 0 and 2, and show
this as the solid line in Fig.~\ref{fig:q}, along with error bars
indicating the local standard deviation $\sigma$.  The median value of
$q'_{\rm IR}$ shows a steady decline with redshift at both 24 and
70~$\mu$m, due to the shape of a typical galaxy SED, which has a
steeper spectrum in the mid-infrared than in the radio.  We define
sources which have a value of $q'_{\rm 24}$ below the local median
value by more than $2\sigma$ to be `AGN-like' -- the typical value of
$\sigma$ for a bin is 0.4, making these sources a factor of $\sim6$
more radio-bright than is typical.  This criterion is similar to that
used by \citet{Ibar08}, although they used $2\sigma$ below the mean
value of $q_{24}$ as their cut-off value.  The median estimator is
more robust to the presence of outliers than the mean; since it is
these outliers which we are trying to identify, the median is a more
appropriate statistic to use.  There are 25 sources out of 510
(5~per~cent) which satisfy this criterion, seven of which are more
than $3\sigma$ away from the local median.  The AGN-like sources are
identified in Fig.~\ref{fig:q}.  For comparison, there are only five
sources that lie above the median value of $q'_{24}$ by more than
$2\sigma$, and none by more than $3\sigma$.  A similar selection for
AGN-like sources could be made in terms of differences from the median
value of $q'_{70}$ -- however, most of the sources that show
significant deviations at 70~$\mu$m have already been identified
through their $q'_{24}$ deviation, and since the resolution of the
24-$\mu$m data is a much better match to the 610-MHz data we choose to
consider deviations from $q'_{24}$ only.

All of the AGN-like sources were visually inspected in the radio
images, in order to look for extended jet/lobe structure.  One source
was potentially the central nucleus of an extended radio source, but
none of the other sources showed evidence for any extended radio
structure.  Fig.~\ref{fig:agn} shows optical, IRAC and MIPS
$50\times50$~arcsec$^{2}$ images centred on each of the AGN-like
sources.  The sources are typically compact, although a few (e.g.\ the
fourth source on the top row of Fig.~\ref{fig:agn}) show signs of
extended emission in the optical images.  Throughout the rest of this
work, the AGN-like sources will be retained in the sample, but
identified as potential contaminants in order to compare them to the
remainder of the population.

%%%%%%%%%%%%%%%%%%%%%%%%%%%%%%%%%%%%%%%%%%%%%%%%%%
\subsection{Star formation rate estimates}
\label{sec:resultssfr}
The SFR estimates for the galaxies in our sample come from the SED
fitting and photometric redshift estimation of
\citet{RowanRobinson08}.  A further method of estimating the SFR,
$\Psi$, from the 1.4-GHz luminosity of galaxies is described by
\citet{Bell03}, calibrated from the total infrared SFR for galaxies
with $L \geq L*$ (defined as having an infrared luminosity $L_{\rm IR}
\sim 2\times10^{10}$~$L_{\odot}$).  Assuming a radio spectral index
$\alpha = 0.8$ \citep[e.g.][]{Condon92}, where we define $\alpha$ such
that the flux density $S = S_{0}\nu^{-\alpha}$, we convert this
relationship to a 610-MHz equivalent:
\begin{equation}
\left(\frac{\Psi}{M_{\odot} {\rm yr}^{-1}}\right) = 2.84\times10^{-22}
\left(\frac{L_{610}}{\rm W~Hz^{-1}}\right) 
\label{eq:Bell1}
\end{equation}
for $L_{\rm 610} > L_{c}$ (where $L_{c} =
3.3\times10^{21}$~W~Hz$^{-1}$ is the luminosity at 610~MHz of a $\sim
L*$ galaxy, with $\Psi \simeq 1$~$M_{\odot}$~yr$^{-1}$), and
\begin{equation}
\left(\frac{\Psi}{M_{\odot} {\rm yr}^{-1}}\right) =
\frac{2.84\times10^{-22}}{0.1 + 0.9(L_{610}/L_{\rm c})^{0.3}}
\left(\frac{L_{610}}{\rm W~Hz^{-1}}\right)  
\label{eq:Bell2}
\end{equation}
for $L_{\rm 610} \leq L_{c}$.

\begin{figure}
  \begin{center}
    \includegraphics[width=0.45\textwidth]{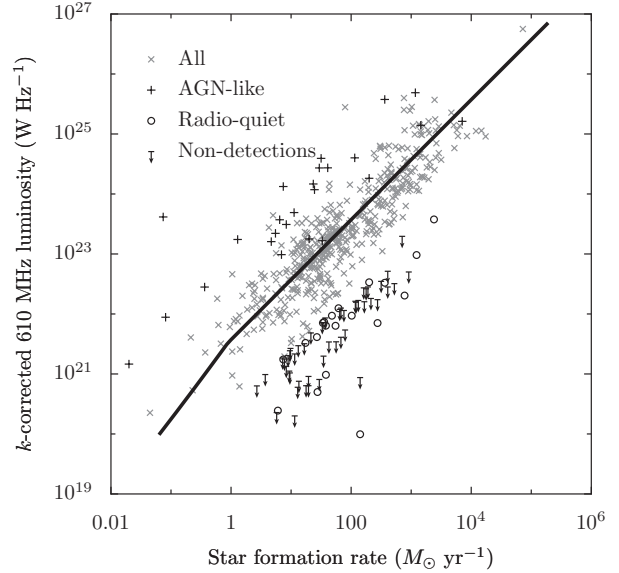}
    \caption{The relationship between $k$-corrected 610-MHz radio
     luminosity and SFR (grey diagonal crosses) and the prediction of
     \citet{Bell03} shifted to 610~MHz assuming $\alpha=0.8$ (solid
     line).  The sources identified as AGN-like are shown as crosses,
     `+', and the sources identified as being radio-quiet (see
     Section~\ref{sec:radioquiet}) are shown as open circles,
     `$\circ$'.  Significant non-detections are indicated by upper
     limits -- see Section~\ref{sec:radioquiet} for further details.}
    \label{fig:lumsfr}
  \end{center}
\end{figure}

Fig.~\ref{fig:lumsfr} shows the relationship between 610-MHz radio
luminosity and SFR.  The radio data were $k$-corrected from an
observation-frame flux density into a rest-frame luminosity using
\begin{equation}
  L_{610} = 4\pi d_{\rm L}^{2} S_{610} (1+z)^{\alpha-1},
  \label{eq:radiokcorr610}
\end{equation}
where we assume that $\alpha=0.8$, and where $d_{\rm L}$ is the
luminosity-distance corresponding to redshift $z$.  The relationship
given by Equations~\ref{eq:Bell1} and \ref{eq:Bell2} is overlaid on
the data, and shows a very good agreement with the overall trend for
sources which are detected in the radio images.  Because the sample
has few galaxies with SFR less than 1~$M_{\odot}$~yr$^{-1}$, we are
unable to confirm the existence of a break at $L < L_{c}$.  The
sources that were previously identified as AGN-like all have a higher
radio luminosity than predicted by Equation~\ref{eq:Bell1} (which
applies to star-forming galaxies only), although some do not deviate
far from the overall trend.

We define the ratio of $L_{610}/\Psi$ to be the `specific radio
luminosity' of a galaxy, and calculate the median value of
log$_{10}(L_{610}/\Psi)$, for all sources with radio detections, and
for sources with SFR $\geq1$~$M_{\odot}$~yr$^{-1}$ which are not
classified as being AGN-like.  The medians and standard deviations are
$21.51\pm0.61$ and $21.48\pm0.55$ respectively, consistent with the
\citet{Bell03} value (shifted to 610~MHz) from Equation~\ref{eq:Bell1}
of 21.55.  In contrast, the radio-luminosity / SFR relationship given
by \citet{Condon90}, which is based around the ratio of the
non-thermal Galactic luminosity at 408~MHz and an estimate of the
Galactic supernova rate, predicts a value of 21.17 at 610~MHz -- also
within one standard deviation of our results, but a less good fit to
the observations.

%%%%%%%%%%%%%%%%%%%%%%%%%%%%%%%%%%%%%%%%%%%%%%%%%%
\subsection{Radio-quiet sources}
\label{sec:radioquiet}
\begin{figure*}
 \centerline{
   \includegraphics[width=0.07\textwidth]{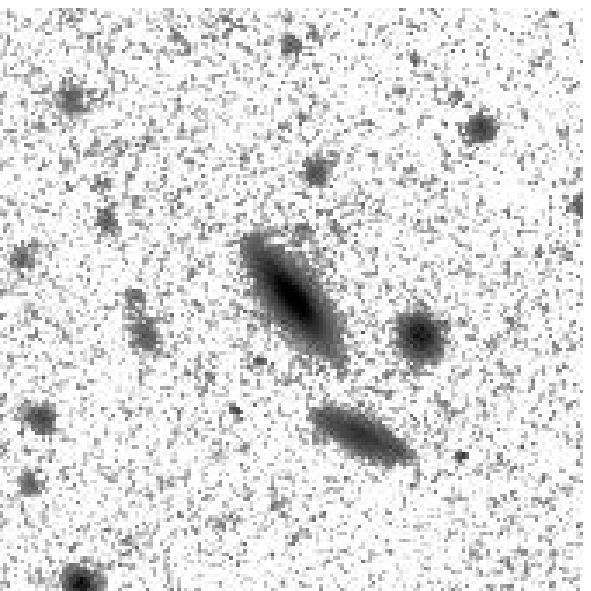}
   \includegraphics[width=0.07\textwidth]{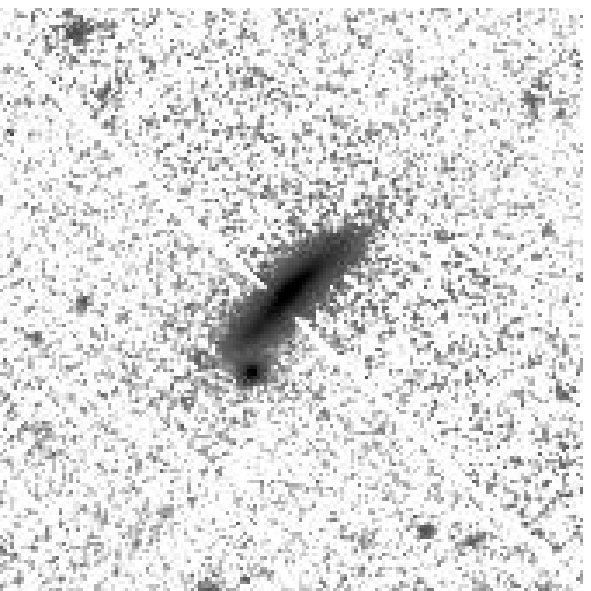}
   \includegraphics[width=0.07\textwidth]{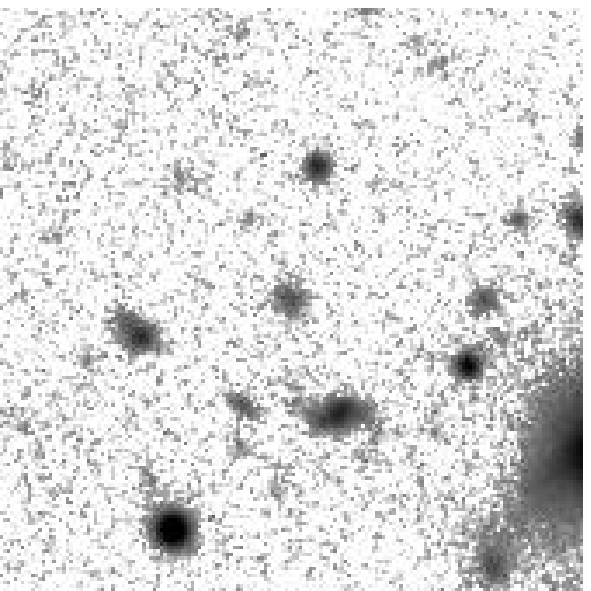}
   \includegraphics[width=0.07\textwidth]{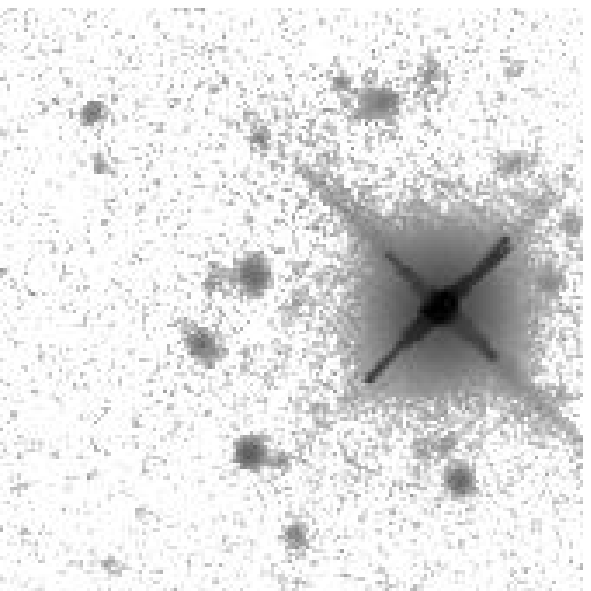}
   \includegraphics[width=0.07\textwidth]{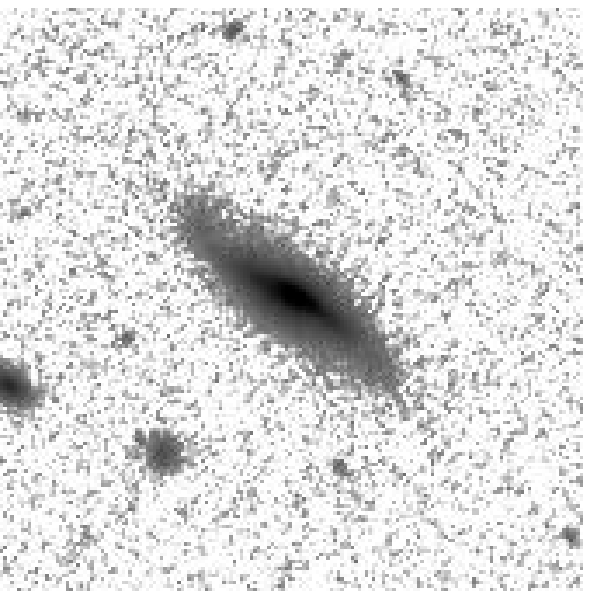}
   \includegraphics[width=0.07\textwidth]{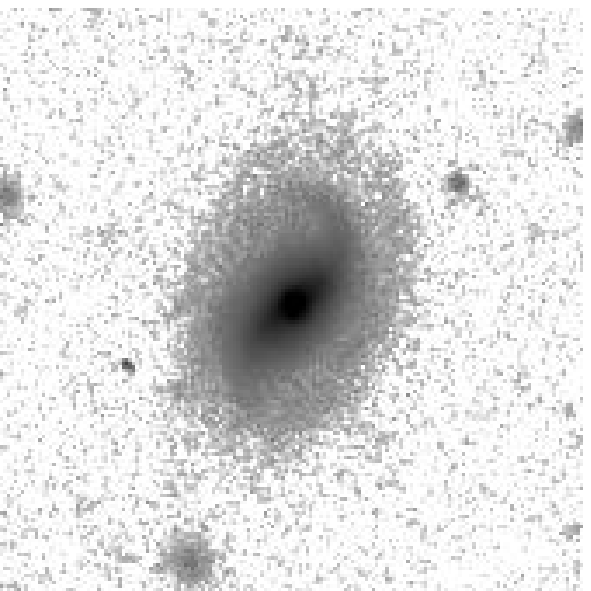}
   \includegraphics[width=0.07\textwidth]{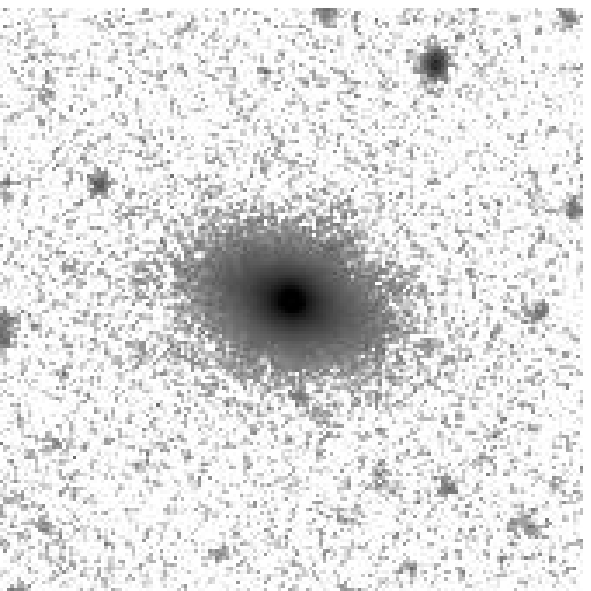}
   \includegraphics[width=0.07\textwidth]{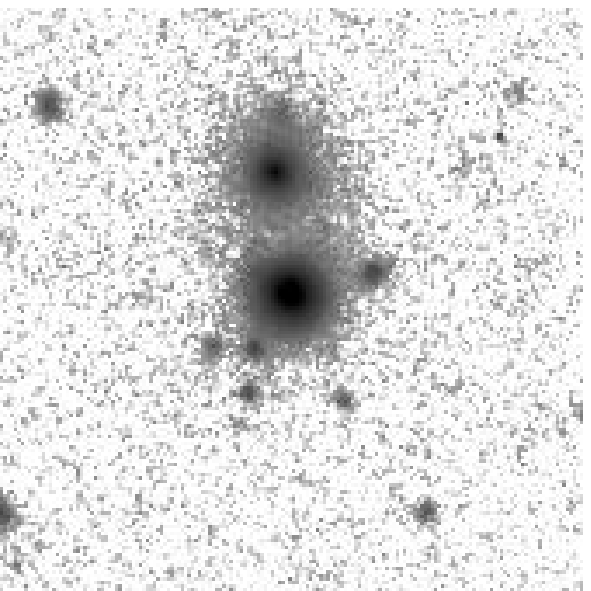}
   \includegraphics[width=0.07\textwidth]{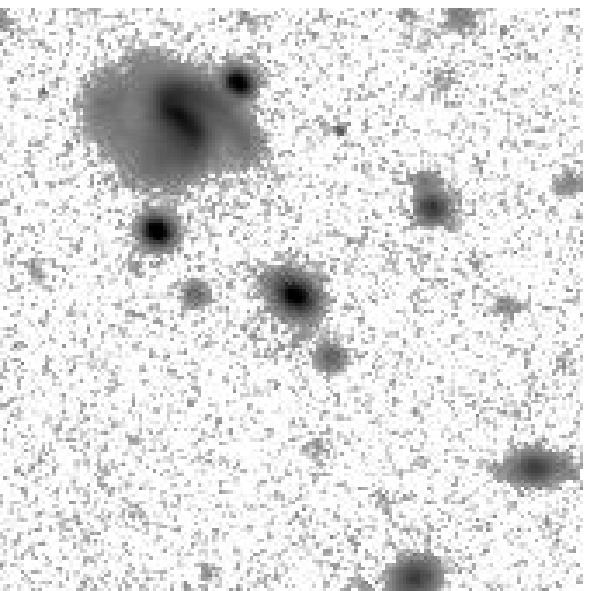}
   \includegraphics[width=0.07\textwidth]{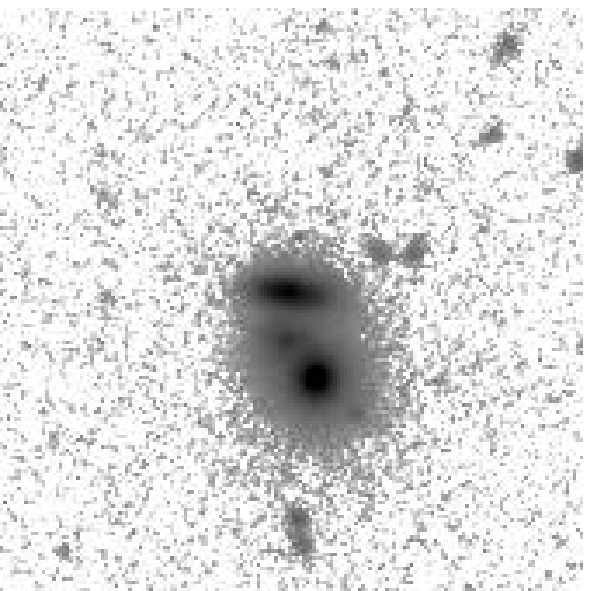}
}
 \centerline{
   \includegraphics[width=0.07\textwidth]{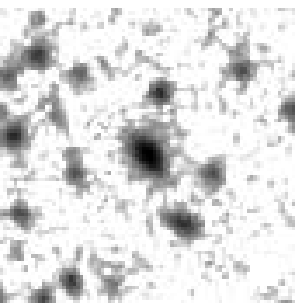}
   \includegraphics[width=0.07\textwidth]{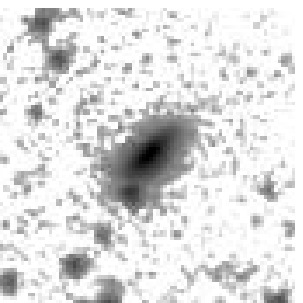}
   \includegraphics[width=0.07\textwidth]{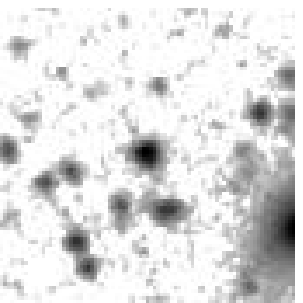}
   \includegraphics[width=0.07\textwidth]{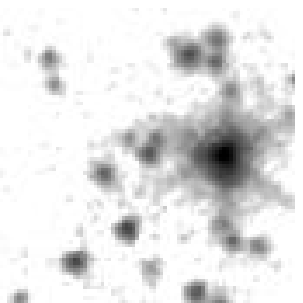}
   \includegraphics[width=0.07\textwidth]{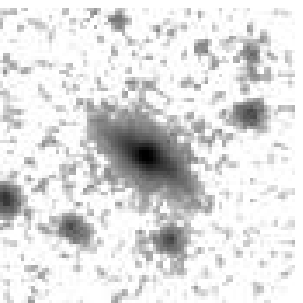}
   \includegraphics[width=0.07\textwidth]{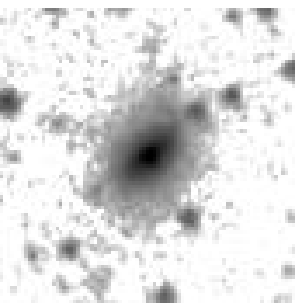}
   \includegraphics[width=0.07\textwidth]{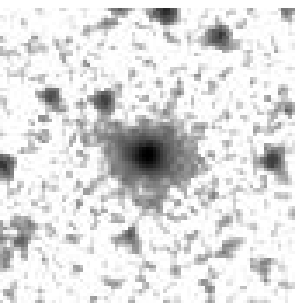}
   \includegraphics[width=0.07\textwidth]{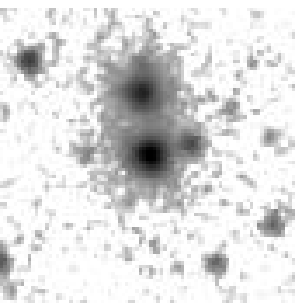}
   \includegraphics[width=0.07\textwidth]{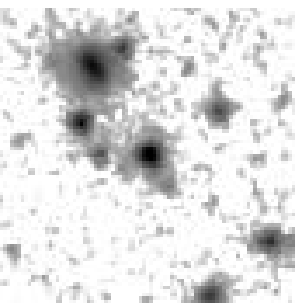}
   \includegraphics[width=0.07\textwidth]{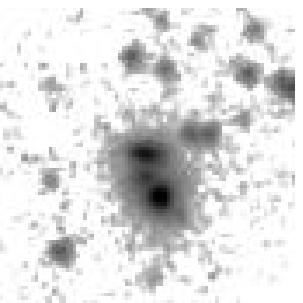}
}
 \centerline{
   \includegraphics[width=0.07\textwidth]{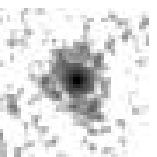}
   \includegraphics[width=0.07\textwidth]{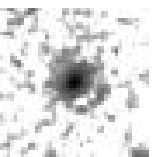}
   \includegraphics[width=0.07\textwidth]{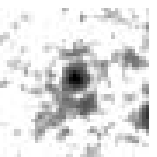}
   \includegraphics[width=0.07\textwidth]{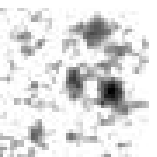}
   \includegraphics[width=0.07\textwidth]{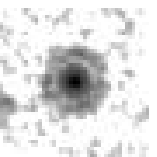}
   \includegraphics[width=0.07\textwidth]{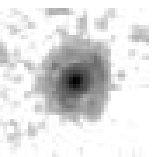}
   \includegraphics[width=0.07\textwidth]{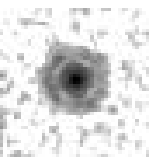}
   \includegraphics[width=0.07\textwidth]{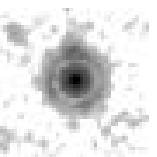}
   \includegraphics[width=0.07\textwidth]{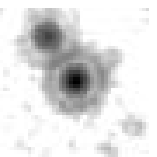}
   \includegraphics[width=0.07\textwidth]{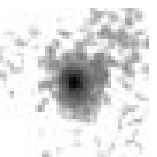}
}
\vskip0.2cm
 \centerline{
   \includegraphics[width=0.07\textwidth]{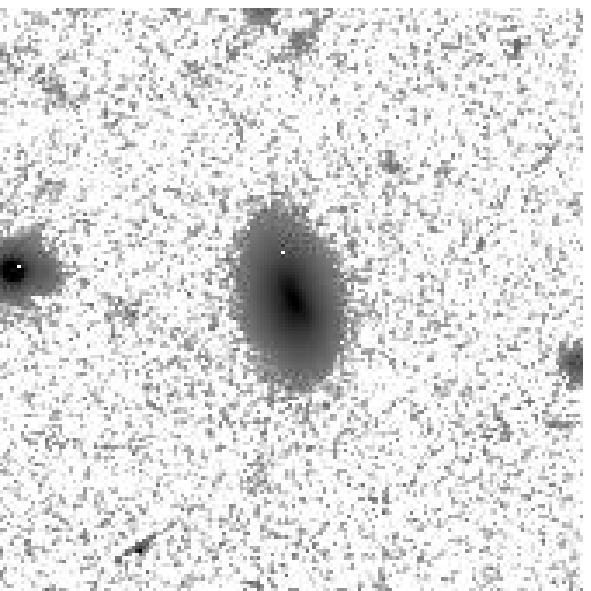}
   \includegraphics[width=0.07\textwidth]{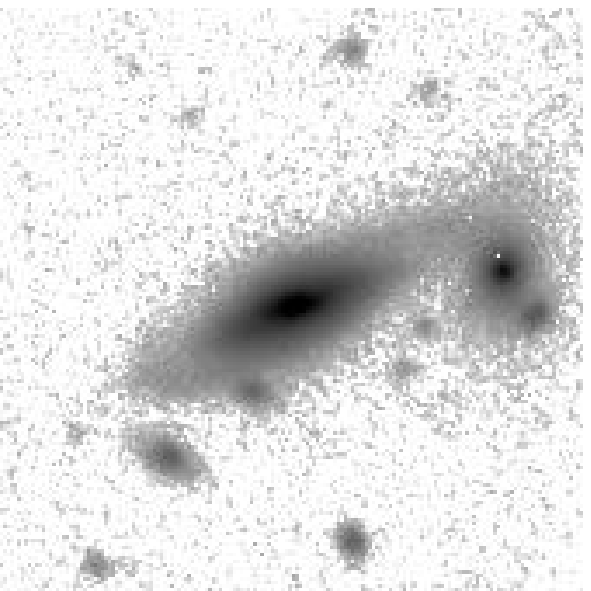}
   \includegraphics[width=0.07\textwidth]{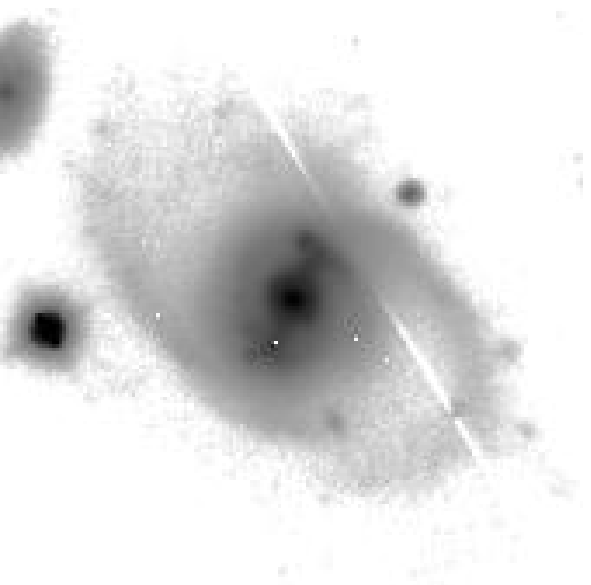}
   \includegraphics[width=0.07\textwidth]{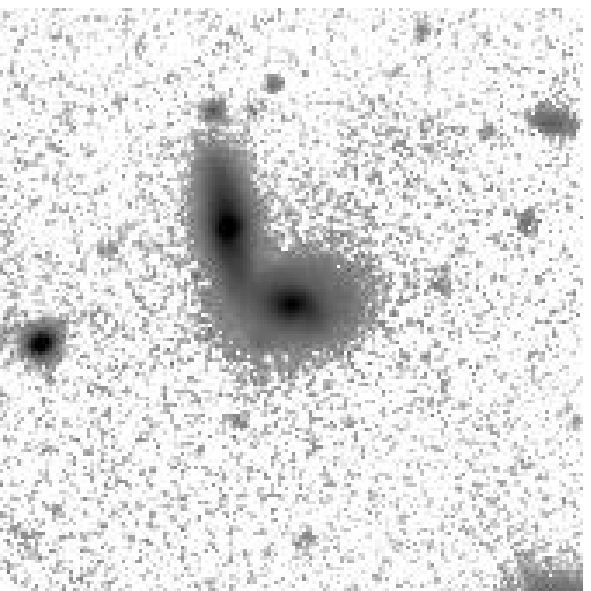}
   \includegraphics[width=0.07\textwidth]{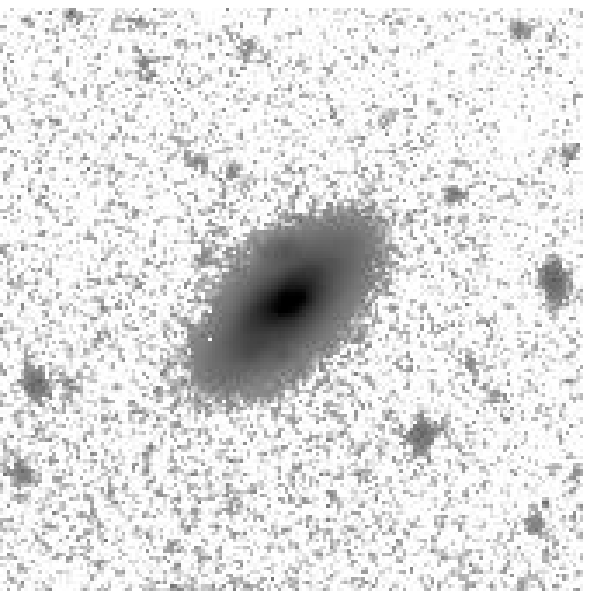}
   \includegraphics[width=0.07\textwidth]{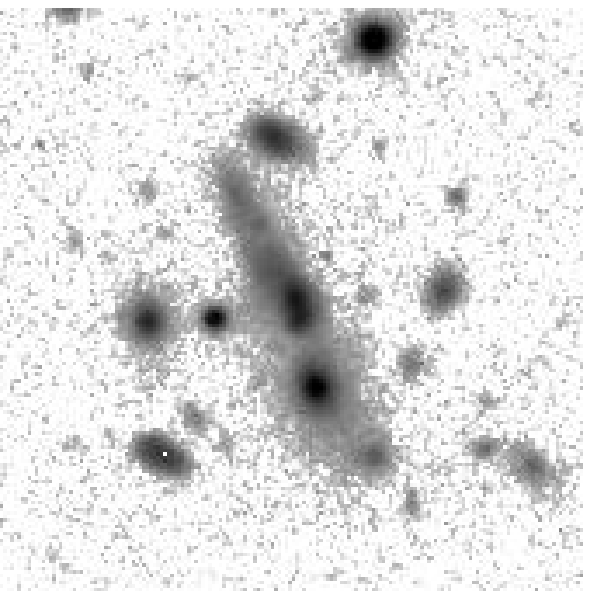}
   \includegraphics[width=0.07\textwidth]{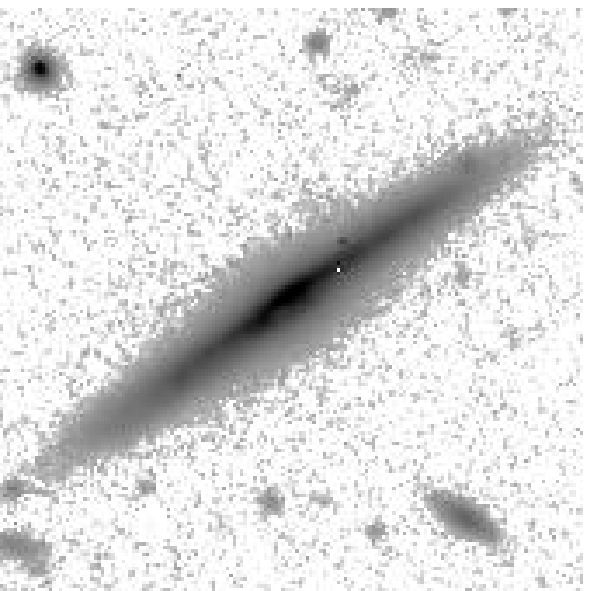}
   \includegraphics[width=0.07\textwidth,height=0.07\textwidth]{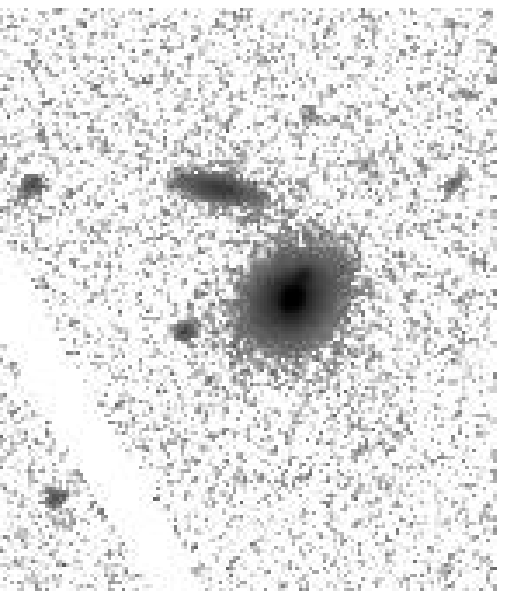}
   \includegraphics[width=0.07\textwidth]{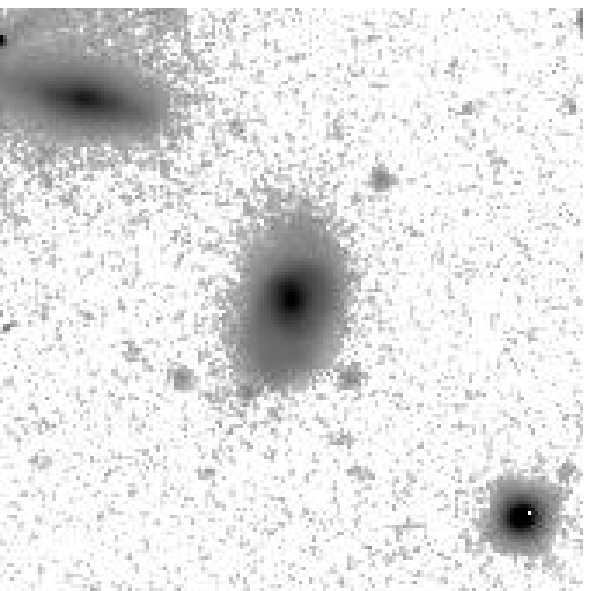}
   \includegraphics[width=0.07\textwidth]{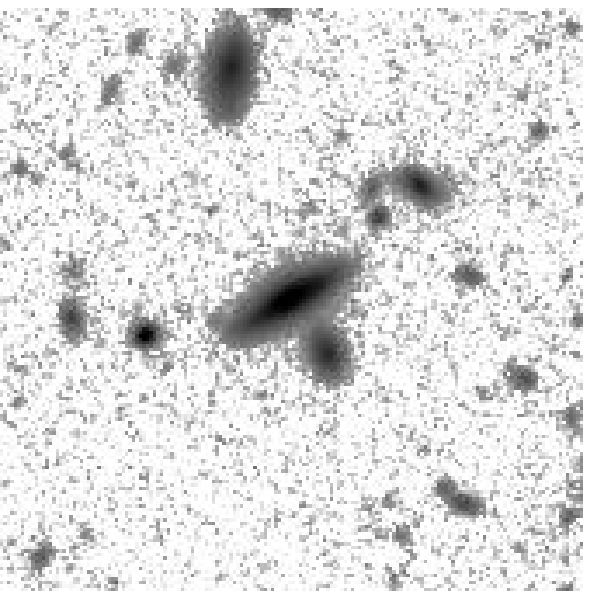}
}
 \centerline{
   \includegraphics[width=0.07\textwidth]{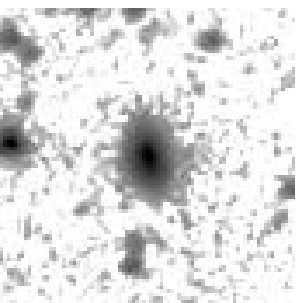}
   \includegraphics[width=0.07\textwidth]{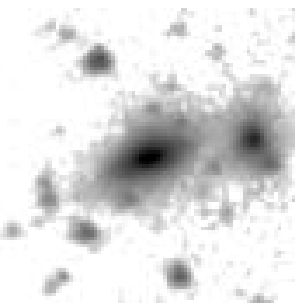}
   \includegraphics[width=0.07\textwidth]{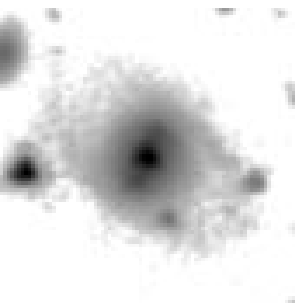}
   \includegraphics[width=0.07\textwidth]{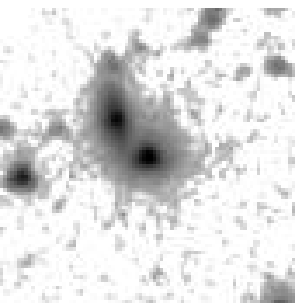}
   \includegraphics[width=0.07\textwidth]{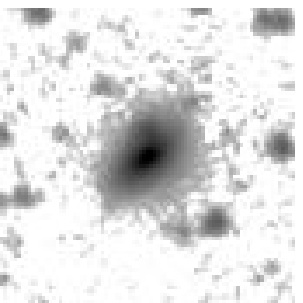}
   \includegraphics[width=0.07\textwidth]{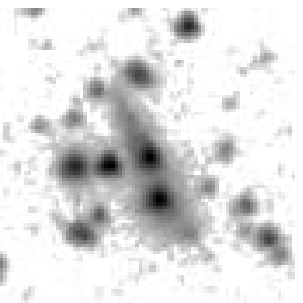}
   \includegraphics[width=0.07\textwidth]{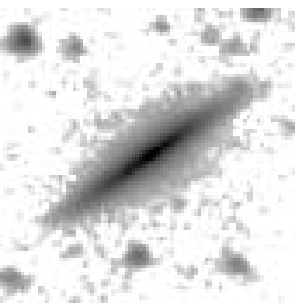}
   \includegraphics[width=0.07\textwidth,height=0.07\textwidth]{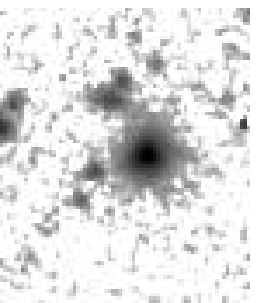}
   \includegraphics[width=0.07\textwidth]{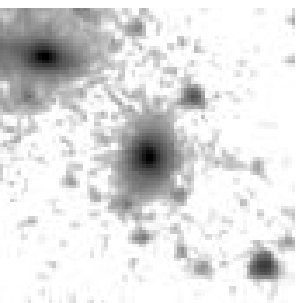}
   \includegraphics[width=0.07\textwidth]{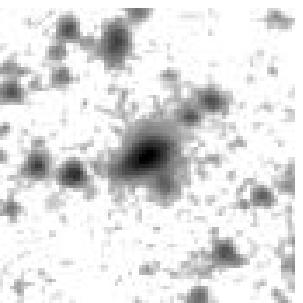}
}
 \centerline{
   \includegraphics[width=0.07\textwidth,angle=90]{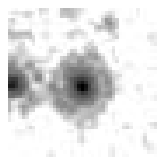}
   \includegraphics[width=0.07\textwidth,angle=90]{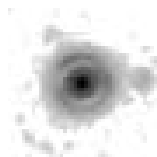}
   \includegraphics[width=0.07\textwidth,angle=90]{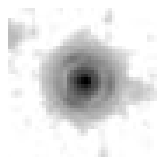}
   \includegraphics[width=0.07\textwidth,angle=90]{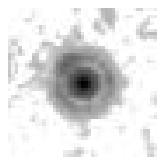}
   \includegraphics[width=0.07\textwidth,angle=90]{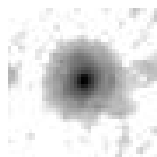}
   \includegraphics[width=0.07\textwidth,angle=90]{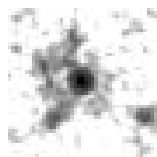}
   \includegraphics[width=0.07\textwidth,angle=90]{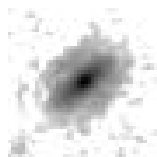}
   \includegraphics[width=0.07\textwidth,angle=90]{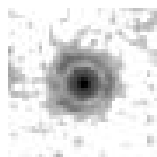}
   \includegraphics[height=0.07\textwidth,angle=90]{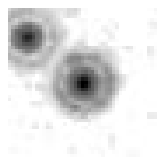}
   \includegraphics[width=0.07\textwidth,angle=90]{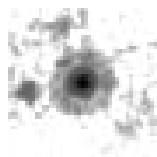}
}
  \caption{Optical and infrared images of the 20 radio-quiet sources,
  retrieved from the NASA/IPAC infrared science archive.  All images
  are $50\times50$~arcsec$^{2}$ in size.  Top row: $r'$-band images.
  Middle row: 3.6-$\mu$m images.  Bottom row: 24-$\mu$m images.}
  \label{fig:rq}
\end{figure*}

There are several sources which are detected in the radio images, but
show significantly lower values of radio luminosity in
Fig.~\ref{fig:lumsfr} than would be expected from their SFR and
Equation~\ref{eq:Bell1}.  We define `radio-quiet' sources as having a
value of log$_{10}(L_{610}/\Psi)$ that is at least $2\sigma$ below the
median value, calculated from non-AGN sources only.  There are 20
sources satisfying this criterion, which are marked as such on
Fig.~\ref{fig:lumsfr}.  The sources are at low redshift (see
Fig.~\ref{fig:q}), with 90~per~cent being at $0 < z < 0.2$.

\begin{figure}
  \begin{center}
    \includegraphics[width=0.45\textwidth]{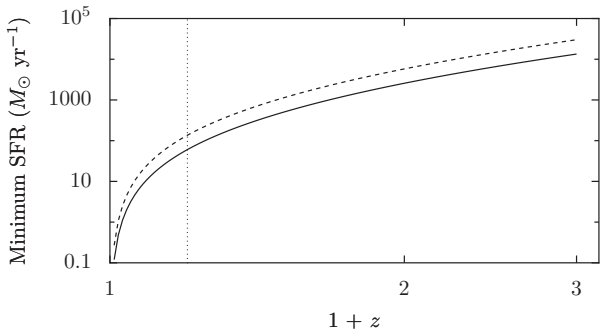}
    \caption{The minimum SFR necessary for sources to be identified as
    radio-quiet, based upon 610-MHz noise levels of
    40~$\mu$Jy~beam$^{-1}$ (solid line) and 90~$\mu$Jy~beam$^{-1}$
    (dashed line).  The vertical dotted line indicates a redshift of
    0.2.}
    \label{fig:sfrlimit}
  \end{center}
\end{figure}

As discussed in Section~\ref{sec:selectionbiases}, the {\it Spitzer}
surveys are more sensitive to star-forming galaxies than the radio
surveys -- of the 5412 infrared-selected sources, only 510 were
detected at 610~MHz.  In order to identify which of the non-detections
are significant, we determine upper limits on 610-MHz flux density for
each source, based upon four times the local noise level, and place
upper limits on the radio luminosity using
Equation~\ref{eq:radiokcorr610}.  We then determine an upper limit on
the value of log$_{10}(L_{610}/\Psi)$ using the
\citet{RowanRobinson08} SFR values -- for the majority of sources,
this value lies above or within the distribution of detected sources
shown in Fig.~\ref{fig:lumsfr} (i.e.\ the radio surveys are not deep
enough to detect the source if it has a `typical' specific radio
luminosity); we consider these non-detections to be insignificant, and
do not discuss them further.  There remain a small number of sources
which have an upper limit on specific radio luminosity that is more
than $2\sigma$ below the value predicted by Equation~\ref{eq:Bell1}
(i.e.\ we would expect to detect these sources if they had a `typical'
specific radio luminosity).  There are 48 significant non-detections,
which are shown on Fig.~\ref{fig:lumsfr}, and appear to be similar to
the `radio-quiet' sources previously identified.

There are a few potential explanations for these sources, which we
consider in turn:
\begin{enumerate}
\item we have significantly under-estimated the radio luminosity for
  these sources (Section~\ref{sec:diffuse});
\item the sources are genuinely radio-quiet, either due to a recent
  burst of star formation, or due to a suppression of the radio
  luminosity within these galaxies (Section~\ref{sec:starburst});
\item the SFR has been significantly over-estimated by
\citet{RowanRobinson08} (Section~\ref{sec:overestsfr}).
\end{enumerate}

%%%%%%%%%%%%%%%%%%%%%%%%%%%%%%%%%%%%%%%%%%%%%%%%%%
\subsubsection{Under-estimation of the radio luminosity}
\label{sec:diffuse}
Any significant under-estimation of the radio flux density would lead
to a lower calculated value of the specific radio luminosity of
galaxies.  We inspected all of the radio-quiet sources in the radio
images in order to test whether the radio counterparts were close to
bright sources, or in noisy parts of the image, but no systematic
effects were apparent.  The sources are mainly at low redshift
(90~per~cent of the radio-quiet galaxies, and 94~per~cent of the
significant non-detections are at $0<z<0.2$), and galaxies are likely
to have a greater angular size in the local Universe than at higher
redshift.  Fig.~\ref{fig:rq} shows that some of the radio-quiet
sources appear to be extended in the optical and infrared images,
suggesting that there may be diffuse radio emission near to or below
the noise level which has not been included into the measured value of
flux density.

The low redshift of these sources can be explained purely through
selection effects -- the SFR that is required for a source to be
identified as either radio-quiet or as a significant non-detection
depends on both the redshift of the source and the local noise level
of the radio image, as illustrated by Fig.~\ref{fig:sfrlimit}.  At
$z=0.2$, a galaxy needs to have a SFR of at least
60~$M_{\odot}$~yr$^{-1}$ to be classified as radio-quiet in the most
sensitive portion of the ELAIS-N1 image, with noise level of
40~$\mu$Jy~beam$^{-1}$, or alternately a SFR greater than
135~$M_{\odot}$~yr$^{-1}$ for a noise level of 90~$\mu$Jy~beam$^{-1}$,
more characteristic of the ELAIS-N2 survey.
Fig.~\ref{fig:distributions}b shows that most of the 24-$\mu$m and
70-$\mu$m selected sources have SFRs below these values, and so could
not be identified as being significantly radio-quiet.

Even if the host galaxy is extended, the majority of star formation in
high SFR galaxies tends to come from a nuclear starburst
\citep[e.g.][]{Kennicutt98}, and the bulk of the radio emission would
not be expected to be spread out over a large area.  In order to test
for extended radio emission we measured the flux density of each of
these sources within a circular aperture centred on their known
location, using the technique described in \citet{Garn09stacking}, and
calculated the ratio of flux density measured within the aperture to
the flux density listed in the catalogue.  For the full sample of
sources, the flux density measured within apertures of varying sizes
was comparable to the catalogued flux density, from which we conclude
that most of the sources have little extended radio emission.  The
AGN-like sources also follow this trend, but the radio-quiet galaxies
show evidence for a slight increase in flux density within apertures
of $\gtrsim15$~arcsec radius, implying the existence of some diffuse
radio emission which has not been catalogued.  The increase is less
than a factor of two; this is not a large enough effect to make these
galaxies significantly radio-quiet.

The conversion between flux density and luminosity relies upon an
assumed radio spectral index of 0.8 for all sources; without
multi-frequency radio data it is not possible to estimate this value
more accurately.  \citet{Garn07} demonstrate that the spectral index
distribution of galaxies detected at both 610~MHz and 1.4~GHz peaks
around $\alpha=0.8$, and various studies
\citep[e.g.][]{Bondi07,Ibar09} have concluded that there is no
significant variation in the mean spectral index of radio sources at
flux densities down to $\sim100$~$\mu$Jy.  Even for an extreme error
in the spectral index of $\sigma_{\alpha}=0.5$, for a galaxy at $z=2$,
the inferred luminosity would not change by more than a factor of two;
again, this is not a large enough effect to explain the
radio-quietness of these sources.

%%%%%%%%%%%%%%%%%%%%%%%%%%%%%%%%%%%%%%%%%%%%%%%%%%
\subsubsection{Physical effects}
\label{sec:starburst}
There are a few potential reasons why galaxies could be radio-quiet
with respect to their SFR; recent star-burst activity, or suppression
of the radio luminosity due to a low confinement of cosmic rays,
potentially caused by low magnetic field strengths.  If a galaxy had
just begun a burst of star formation activity that has enhanced the
SFR, but not yet enhanced the supernova rate, it would appear to be
radio-quiet -- since the supernova rate lags the SFR by $\sim30$~Myr
(the approximate lifetime of 8~$M_{\odot}$ stars), this would mean
that sources were being observed very close to the beginning of the
starburst period.  Some of the sources (e.g.\ the second source in row
two of Fig.~\ref{fig:rq}) show signs of potential on-going mergers in
their optical images, which would be an indication of triggered star
formation, although this is not the case for all sources.  A full
analysis of the possibility of mergers in the sample is beyond the
scope of this paper, although it seems unlikely that all of the
radio-quiet sources in our sample could be undergoing merger activity.

The best-studied starburst galaxy in the local Universe is Arp~220, at
$z = 0.018126$ \citep{deVaucouleurs91} with SFR of
240~$M_{\odot}$~yr$^{-1}$, calculated from its FIR luminosity
\citep{Sanders03} and the relationship given in \citet{Kennicutt98}.
It has a 1.4-GHz flux density of 0.326~Jy \citep{Condon02} and a
365-MHz flux density of 0.435~mJy \citep{Douglas96}, giving it a
spectral index $\alpha \simeq 0.2$.  This equates to a luminosity of
$2.5\times10^{23}$~W~Hz$^{-1}$ at 610~MHz, and a ratio of ${\rm
log}_{10}\left(L_{610}/\Psi\right)$ of $\sim21$, within $1\sigma$ of
the median value found in Section~\ref{sec:resultssfr}, thus Arp~220
would not have been identified as being radio-quiet through this
diagnostic.  However, this should not be surprising -- Arp~220 has a
large number of compact radio sources visible
\citep[e.g.][]{Smith98,Lonsdale06} which are thought to be radio
supernovae formed due to the starburst activity -- this system has
therefore already passed the phase at which it could be identified as
radio-quiet.  The findings of \citet{Wilson06} agree with this
interpretation -- there are a number of star clusters with ages up to
around 500~Myr within Arp~220, suggesting that starburst activity
started a long time ago.  Any starburst galaxies identified through
through this diagnostic would have to be significantly earlier on in
their starburst phase than Arp~220.

\citet{Carilli08} have identified evidence for a possible decrease in
the conversion factor between radio luminosity and SFR for high
redshift ($z\sim3$) Lyman break galaxies, based upon a stacking
analysis of sources in the COSMOS field \citep[although see][for a
discussion of the biases present in radio stacking
experiments]{Garn09stacking}.  Their favoured explanation of increased
relativistic electron cooling due to inverse Compton scattering off
the cosmic microwave background (CMB) would not apply to our
low-redshift radio-quiet galaxies, since the energy density in the CMB
increases with redshift as $(1+z)^{4}$ and is comparatively
unimportant in the local Universe.  A potential explanation for
low-redshift radio-quiet sources would be that they have lower
magnetic field strengths than is typical, leading to a decreased
electron confinement \citep[for the optically-thin scenario discussed
by][]{Chi90} and a lower fraction of the total energy in cosmic rays
being radiated within a galaxy compared with a source having a greater
$B$-field.  However, the radio-quiet sources follow the same
infrared-radio correlation as the main sample (shown by
Fig.~\ref{fig:q}) and a decrease in radio luminosity would therefore
also require a corresponding decrease in the infrared luminosity; if
this were the case, the measured SFR (which was calculated from the
infrared luminosity) should also decrease, and sources would not be
identified as being radio-quiet.

%%%%%%%%%%%%%%%%%%%%%%%%%%%%%%%%%%%%%%%%%%%%%%%%%%
\subsubsection{Over-estimation of the star formation rate}
\label{sec:overestsfr}
The SFR for these galaxies has been estimated using a complex method,
described in \citet{RowanRobinson08}:
\begin{enumerate}
\item optical and IRAC fluxes are used to find the best-fitting
  template spectrum for each galaxy, from a selection of six
  templates, and photometric redshifts are calculated;
\item sources with an `infrared excess' at $\lambda \geq$ 8~$\mu$m
  have their bolometric infrared luminosity $L_{\rm IR}$ estimated
  from the long wavelength data;
\item 60-$\mu$m luminosities, $L_{60}$, are calculated from $L_{\rm
  IR}$, using bolometric correction factors of 3.48, 1.67 and 1.43 for
  cirrus, M82 and Arp~220 galaxy templates respectively.
\item SFRs are then estimated from the 60-$\mu$m luminosity, using
  the conversion factor of
\begin{equation}
  \left(\frac{\Psi}{M_{\odot}{\rm yr}^{-1}}\right) = 2.2 \epsilon^{-1}
  10^{-10} \left(\frac{L_{60}}{L_{\odot}}\right),
\end{equation}
where $\epsilon$ describes the fraction of UV light absorbed by dust,
and is taken to be 2/3.  
\end{enumerate}

\citet{RowanRobinson08} estimate an uncertainty in the total infrared
luminosity at $z=0.2$ (the approximate redshift of the radio-quiet
sources) of 0.2~dex -- this is too small to be responsible for the
deviations which are seen.  A more plausible scenario is for source
confusion to play an important role -- the resolution at 70~$\mu$m is
only $\sim18$~arcsec, much poorer than in the optical, IRAC or
24-$\mu$m bands.  If there are two distinct sources present in the
optical images, only one of which has been assigned a far-infrared
counterpart (with the combined infrared flux from both) then that
source will have its SFR overestimated in the photometric redshift
catalogue, while the other source will not appear in the sample due to
its lack of an infrared detection.  Inspection of Fig.~\ref{fig:rq}
shows several sources where confusion could potentially be having an
effect on the SFR estimation.  However, it could only plausibly
increase the flux density by a factor of $\sim2$, and is not likely to
lead to an order of magnitude discrepancy.

The conversion between $L_{\rm IR}$ and $L_{60}$ has uncertainties in
it, but again there is only a factor of $\sim2$ difference between the
three templates that are used.  Likewise, the parameter $\epsilon$
used to calculate the amount of young star formation could be in error
for galaxies with a significant old stellar population
\citep{Bell03,RowanRobinson03} which would lead to an over-estimation
of the SFR, but neither of these factors could explain such a large
difference in SFR for these sources.  

\begin{figure}
  \begin{center}
    \includegraphics[width=0.45\textwidth]{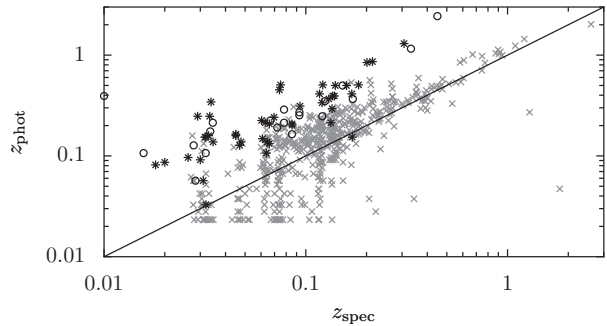}
    \caption{Comparison of spectroscopic and photometric redshifts for
    all galaxies in our 24-$\mu$m and 70-$\mu$m selected sample with
    spectroscopy (grey diagonal crosses), the radio-quiet sources
    (open circles) and the significant non-detections (stars).  The
    1:1 line is indicated for reference.}
    \label{fig:redshiftcomparison}
  \end{center}
\end{figure}

The fact that the radio-quiet sources follow the same infrared / radio
correlation as the main sample, but yet show a disagreement between
their radio luminosity and infrared-derived SFR implies that the
template-fitting carried out by \citet{RowanRobinson08} to estimate
photometric redshifts and total infrared luminosity may have been
unsuccessful for a small number of sources -- if this were the case,
then the values of SFR derived from this process would be inaccurate.
In order to test for this, in Fig.~\ref{fig:redshiftcomparison} we
compare the photometric and spectroscopic redshifts for all of the
galaxies which have available spectroscopy, and were selected at 24
and 70~$\mu$m.  All 20 of the radio-quiet sources, and 46 of the 48
significant non-detections have spectroscopic redshifts, in contrast
to the fraction of spectroscopic redshifts available for the whole
sample (576/5412; 11~per~cent).  The \citet{RowanRobinson08} criterion
for defining a catastrophic redshift error is $\Delta{\rm log}_{10}(1
+ z) = \pm 0.06$, or 15~per~cent; of the 66 radio-quiet or significant
non-detected sources with spectroscopic redshifts, 30 (45~per~cent)
were outside this threshold.  In contrast, only four of the sources
which were detected at 610-MHz and not found to be radio-quiet were
classified as having a catastrophic redshift error.  We exclude these
34 sources from the remainder of the analysis, as the SFRs calculated
using inaccurate templates will be unreliable.

\begin{figure}
  \centerline{\subfigure[]{
    \includegraphics[width=0.45\textwidth]{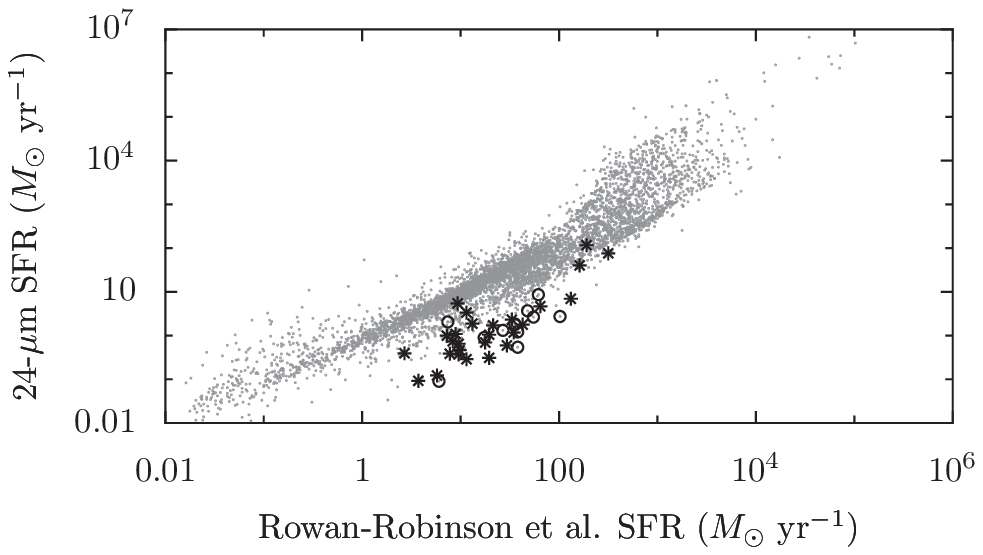}}}
  \centerline{\subfigure[]{
    \includegraphics[width=0.45\textwidth]{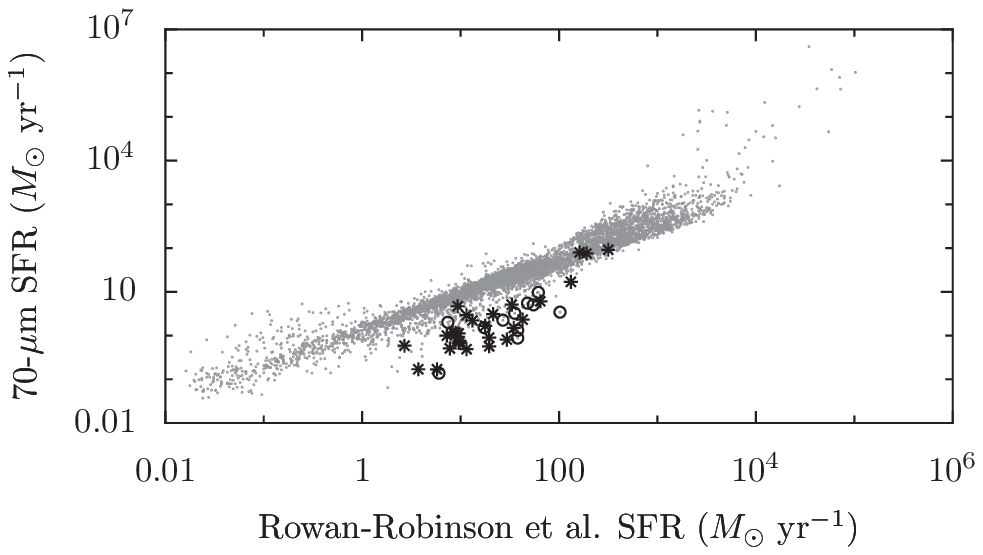}}}
  \caption{The relationship between SFRs estimated by the
  template-fitting of \citet{RowanRobinson08}, and SFRs estimated from
  the relationships given in \citet{Rieke09}, using (a) 24-$\mu$m flux
  densities; (b) 70-$\mu$m flux densities.  Galaxies in the 24-$\mu$m
  and 70-$\mu$m-selected sample are shown as grey dots, with the
  radio-quiet sources indicated by open circles, and the significant
  non-detections shown as stars.}
  \label{fig:sfr2470RR}
\end{figure}

In order to test whether the SFRs for the remainder of the radio-quiet
and significant non-detected sources are reliable, we use
relationships provided by \citet{Rieke09} to calculate SFRs directly
from monochromatic flux densities measured by MIPS at either 24 or
70~$\mu$m.  While SFRs based around a single flux measurement are
likely to be less accurate than SFRs based around bolometric
luminosities, \citet{Rieke09} believe that they should be accurate to
within about 0.2~dex, and these estimates can be used as an
independent verification of the accuracy of the
\citet{RowanRobinson08} SFRs.  In Fig.~\ref{fig:sfr2470RR} we compare
these three SFR estimates for our 24-$\mu$m and 70-$\mu$m selected
sample.  There is some scatter in the data, but it is clear that the
galaxies which are radio-quiet or significant non-detections are
typically outliers, with their monochromatic SFRs being smaller than
the estimates from \citet{RowanRobinson08}.

We favour an over-estimation of the SFR in a small fraction of our
sample for explaining the apparent low specific radio luminosity of a
few of the sources, although we can not rule out a contribution from
the other possibilities.  We note that if SFRs calculated from either
monochromatic 24-$\mu$m or 70-$\mu$m flux densities were used instead
of the \citet{RowanRobinson08} values for these sources, only two of
the significant non-detections would remain significant, and none of
the radio-quiet sources would still be classified as such.  For
completeness the radio-quiet sources and sources with a significant
non-detection, which have not been rejected for poor template-fitting,
will be retained on future plots for comparison with the rest of the
population.

%%%%%%%%%%%%%%%%%%%%%%%%%%%%%%%%%%%%%%%%%%%%%%%%%%
\section{Discussion}
\label{sec:discussion}
\citet{Bell03} presented a relationship between the SFR and 1.4-GHz
non-thermal radio luminosity of galaxies in the local Universe.  This
was calibrated from local galaxies, assuming a direct proportionality
between non-thermal radio emission and SFR for galaxies with $L_{\rm
IR} > L*$.  By constructing a sample of galaxies within the northern
SWIRE regions, we have demonstrated that this relationship can be
successfully applied to galaxies that are detected at 610~MHz.  We
compared the 610-MHz radio luminosity ($k$-corrected assuming a
synchrotron radio spectrum with $\alpha=0.8$) to the SFR in order to
calculate a parameter, $L_{610}/\Psi$, which we call the `specific
radio luminosity' of star-forming galaxies.  This value shows good
agreement with the relationship given by \citet{Bell03}, and a less
good match to the more commonly-used relationship of \citet{Condon90}.
We find that approximately 4~per~cent of the detected sources do not
follow this relationship, and note the existence of several more
galaxies that would be expected to be detected at 610~MHz, but also
appear radio-quiet.  Of the total number of 24-$\mu$m and 70-$\mu$m
selected galaxies which could potentially be detected at 610~MHz, we
find 12~per~cent (68/558) are significantly more radio-quiet than the
\citet{Bell03} relationship would suggest from their SFRs.  We suggest
a number of potential explanations for these sources, and favour an
inappropriate conversion between infrared flux densities and SFRs for
a small number of galaxies, due to an uncertainty in the photometric
redshift calculations, rather than a real physical effect.

Various studies appear to indicate that there has been little
evolution in the infrared / radio correlation at high redshift
\citep[e.g.][]{Garrett02,Appleton04,Ibar08}, suggesting that there has
been little change in the physical processes which link star
formation, thermal dust emission and radio luminosity, although these
studies work with sources selected in the radio and are therefore
biased against radio-quiet galaxies.  In this work we do not directly
test for variations in the luminosity correlation with redshift.
However, the SFR estimates provided by \citet{RowanRobinson08} allow a
probe of any evolution in the physical link between SFR and radio
synchrotron luminosity at high redshift, which has the advantage over
studies which look at either the luminosity or flux density
correlation in that we are working with one physical parameter (the
SFR) and one potentially variable process (synchrotron radiation),
rather than two processes (thermal dust emission at a given wavelength
and synchrotron radiation) that could both vary with redshift.

\begin{figure}
  \begin{center}
    \includegraphics[width=0.45\textwidth]{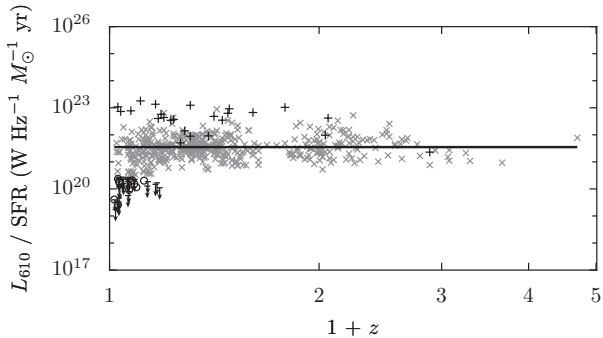}
    \caption{The relationship between 610-MHz radio luminosity and
    SFR, as a function of redshift -- symbols are the same as for
    Fig.~\ref{fig:lumsfr}.  The local relationship given by
    \citet{Bell03} is indicated by the solid line -- there is no
    evidence for variation in the radio luminosity / SFR relationship
    out to $z=2$, for radio-selected star-forming galaxies.}
    \label{fig:swirelumsfrz}
  \end{center}
\end{figure}

In Fig.~\ref{fig:swirelumsfrz} we show the specific radio luminosity
of the galaxies in our sample, as a function of redshift.  Any
variation seen in specific radio luminosity with $z$ would suggest a
change in one or more of the properties that controls the synchrotron
radiation from star-forming galaxies -- specifically, the galactic
magnetic field strength, or the amount of confinement of the
high-energy electrons.  However, no such variation is seen -- the
majority of galaxies which are detected in the radio follow the global
relationship given by \citet{Bell03} out to a redshift of at least 2
(a look-back time of 10.3~Gyr) and potentially to $z\sim3$.  The
AGN-like sources discussed in Section~\ref{sec:resultsq} all have a
high specific radio luminosity (which is to be expected if some of
their radio luminosity is not related to star formation activity), but
their specific radio luminosity shows no evidence for redshift
evolution.  The radio-quiet sources discussed in
Section~\ref{sec:radioquiet} all lie at low redshift and do not affect
this conclusion, although we caution that due to selection effects, we
are unable to test for the existence or the number of radio-quiet
sources which may exist at redshifts beyond 0.2.

As discussed in Section~\ref{sec:introduction}, there is much interest
in trying to calculate the star formation history of the Universe.
Radio observations at $\sim1$~GHz are unaffected by dust, and should
allow a reliable estimation of the variation in SFRD with $z$, without
the uncertainties that are found at optical and infrared wavelengths.
There have been a few attempts to carry out this calculation
\citep[e.g.][]{Mobasher99,Haarsma00,Seymour08,Smolcic09SFR} but all of
these rely upon the untested assumption that the local relationship
between radio luminosity and SFR can be applied successfully at higher
redshift.  The work presented here demonstrates that this assumption
is valid -- the radio luminosity of the majority of radio-selected
star-forming galaxies does track the SFR out to at least $z=2$ (with
some potential evidence of an extension out to $z\sim3$ and beyond;
see Fig.~\ref{fig:swirelumsfrz}).

\begin{figure}
  \begin{center}
    \includegraphics[width=0.45\textwidth]{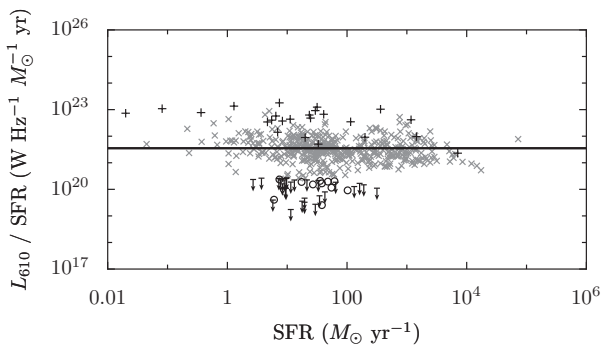}
    \caption{The relationship between 610-MHz radio
    luminosity and infrared-derived SFR, as a function of SFR --
    symbols are the same as for Fig.~\ref{fig:lumsfr}.  The
    relationship given by \citet{Bell03} is indicated by the solid
    line.}
    \label{fig:lumsfrsfr}
  \end{center}
\end{figure}

Equally importantly, the relationship between luminosity and SFR has
been implicitly assumed to be the same for galaxies with all values of
SFR.  Fig.~\ref{fig:lumsfrsfr} shows that, at least over the range 1
-- 10$^{4}$~$M_{\odot}$~yr$^{-1}$, no significant variation in this
relationship is seen.  Galaxies undergoing extreme starbursts, with
SFRs up to $10^{4}$~$M_{\odot}$~yr$^{-1}$, and galaxies undergoing
much lower star-formation activity with SFR of 1~$M_{\odot}$~yr$^{-1}$
show the same relationship between their synchrotron emission and SFR.
The deviation predicted by \citet{Bell03} below
$\sim1$~$M_{\odot}$~yr$^{-1}$ cannot be tested in this work, due to a
lack of low-luminosity sources in the sample.

We have demonstrated that the \citet{Bell03} relationship between
1.4-GHz luminosity and SFR (for $L\geq L*$ galaxies) can be used
successfully on galaxies at $0<z<2$, although it is necessary to first
apply a $k$-correction to the observed flux density of a galaxy in
order to convert to a rest-frame 1.4-GHz luminosity.
Equation~\ref{eq:sfrz} gives a generalised version of this
relationship, obtained through combining the \citet{Bell03} equation
with Equation~\ref{eq:radiokcorr610},
\begin{equation}
  \left(\frac{\Psi}{M_{\odot}~{\rm yr}^{-1}}\right) =
  \frac{0.066}{1+z}\left(\frac{d_{\rm L}(z)}{\rm Mpc}\right)^{2}
  \left(\frac{(1+z)\nu}{\rm 1.4~GHz}\right)^{\alpha}
  \left(\frac{S_{\nu}}{\rm Jy}\right).
\label{eq:sfrz}
\end{equation}
This equation can be used to predict the SFR of a galaxy directly from
observational radio data.

%%%%%%%%%%%%%%%%%%%%%%%%%%%%%%%%%%%%%%%%%%%%%%%%%%
\section{Conclusions}
We have presented new 610-MHz GMRT observations of the ELAIS-N2
region, covering 6~deg$^{2}$ with 13 pointings.  The typical noise
level of each pointing is $\sim$90~$\mu$Jy~beam$^{-1}$ before primary
beam correction.  These radio observations have been combined with our
previous surveys of the ELAIS-N1 and Lockman Hole regions, along with
optical and infrared surveys, in order to create a sample of 510
galaxies in the three northern SWIRE fields, which have photometric
redshifts and SFR estimates from \citet{RowanRobinson08}, infrared
detections at 24 and 70~$\mu$m, and radio detections at 610~MHz.

We have used the logarithmic ratio of infrared to radio flux
densities, $q'_{\rm IR}$, to discriminate between sources which are
powered by star formation processes, and sources which may have a
significant fraction of their radio emission resulting from AGN
activity.  By comparing the rest-frame 610-MHz radio luminosity to the
\citet{RowanRobinson08} SFR, we have identified a further set of
sources which appear to be radio-quiet; we have demonstrated that
these are likely to be galaxies where the template-fitting has been
unsuccessful, leading to an over-estimation of the photometric
redshifts and SFRs, although we discuss a number of real physical
effects which could also be contributing to a reduction in the radio
luminosity, such as the recent onset of starburst activity.

We have compared the \citet{Condon90} and \citet{Bell03} relationships
between SFR and radio luminosity, and found that the \citet{Bell03}
prescription is a better fit to our sample.  By considering the ratio
of radio luminosity and SFR, the `specific radio luminosity', we have
tested the validity of two of the assumptions that are commonly made
when calculating the SFR history of the Universe from deep radio
observations, namely that a relationship calibrated from local
galaxies is applicable to galaxies at much higher redshift, and
undergoing much greater starburst activity than seen in the local
Universe.  We find no redshift dependence for the specific radio
luminosity, suggesting that there has been little change in the
physical processes linking synchrotron radiation and star formation in
star-forming galaxies over the redshift range $0<z<2$.  No variation
in the specific radio luminosity of galaxies with their SFR was found
for a SFR in the range of 1 to $10^{4}$~$M_{\odot}$~yr$^{-1}$.  This
implies that the link between SFR and synchrotron luminosity is the
same for massive starburst galaxies as it is for the more quiescent
galaxies seen in the local Universe.  We conclude that the
\citet{Bell03} relationship between radio luminosity and SFR,
calibrated from local galaxies, can successfully be applied to
high-redshift, high-SFR galaxies, and present a generalised equation
to link the SFR of a galaxy to the observed flux density, redshift and
radio spectral index.

%%%%%%%%%%%%%%%%%%%%%%%%%%%%%%%%%%%%%%%%%%%%%%%%%%
\section*{Acknowledgements}
TG thanks Dominic Ford for useful discussions, and the UK STFC for a
Studentship.  We thank the anonymous referee for helpful suggestions,
and the staff of the GMRT who have made these observations possible.
The GMRT is operated by the National Centre for Radio Astrophysics of
the Tata Institute of Fundamental Research, India.

%%%%%%%%%%%%%%%%%%%%%%%%%%%%%%%%%%%%%%%%%%%%%%%%%%
\setlength{\labelwidth}{0pt}
\bibliography{References}
\bibliographystyle{mn2e}
\label{lastpage}

\end{document}